\documentclass[aps,preprint]{revtex4-1}
\usepackage{amsmath,graphicx,bbm,mathrsfs,amssymb,pst-all,bm,color}
\usepackage{subfigure}

\begin{document}

\title{Efficient quantum algorithm for solving structured problems via
multi-step quantum computation}
\author{Hefeng Wang$^1$}
\email{wanghf@mail.xjtu.edu.cn}
\author{Sixia Yu$^2$}
\email{yusixia@ustc.edu.cn}
\author{Hua Xiang$^3$}
\email{hxiang@whu.edu.cn}

\affiliation{$^{1}$Department of Applied Physics, School of Physics, Xi'an
Jiaotong University and Shaanxi Province Key Laboratory of Quantum
Information and Quantum Optoelectronic Devices, Xi'an, 710049, China}
\affiliation{$^{2}$Hefei National Laboratory for Physical Sciences at
Microscale and Department of Modern Physics, University of Science and
Technology of China, Hefei, Anhui 230026, China}
\affiliation{$^{3}$School of Mathematics and Statistics, Wuhan University, Wuhan,
430072, China}

\begin{abstract}
In classical computation, a problem can be solved in multiple steps where
calculated results of each step can be copied and used repeatedly. While in
quantum computation, it is difficult to realize a similar multi-step
computation process because the no-cloning theorem forbids making copies of
an unknown quantum state perfectly. We find a method based on quantum
resonant transition to protect and reuse an unknown quantum state that
encodes calculated results of an intermediate step without making copies of
the state, and present a quantum algorithm that solves a problem via a
multi-step quantum computation process. This algorithm can achieve an
exponential speedup over classical algorithms in solving a type of
structured search problems.
\end{abstract}

\maketitle

\section{Introduction}

Solving a problem on a quantum computer can be transformed to finding the ground state of a problem Hamiltonian that encodes the solution to the problem. The phase estimation algorithm~(PEA)~\cite{kitaev, abrams} projects an initial state onto the ground state of the problem Hamiltonian with probability proportional to the square of the
overlap between them. However, it is difficult to find a good initial state
for a complicated system. By using amplitude amplification, quantum
algorithms can achieve quadratic speed-up over classical algorithms in
preparing the ground state of a quantum many-body system~\cite{poulin1}. In
adiabatic quantum computing~(AQC)~\cite{farhi1}, the system is evolved
adiabatically from the ground state of an initial Hamiltonian to that of the
problem Hamiltonian. The efficiency of AQC depends on the minimum energy gap
between the ground and the first excited states of the adiabatic
Hamiltonian, which is difficult to estimate in most cases. Quantum Zeno
effect~\cite{childs, poulin} can be used to keep a quantum computer near
ground state of a smoothly varying Hamiltonian by performing frequent
measurements, and has the same efficiency as AQC.

The structure of a problem is the key for whether it can be solved
efficiently or not on a quantum computer. In Refs.~\cite{struct, structAQC},
a nested search algorithm was proposed for problems that can be divided into
two~(or more) levels described by a set of primary and secondary variables,
respectively. It works by nesting one quantum search within another, and
performing quantum search at a selected level among partial solutions to
narrow subsequent search over their descendants. The complete solution is
constructed through a tree of partial solutions at different levels. This
algorithm achieves quadratic speedup over the corresponding classical nesting algorithms, and can be faster than the usual Grover bound for unstructured search. In general, the constraints of a problem contain variables that are coupled to each other, the variables may be divided into only a few sets, thus the search space is still exponentially large.
\begin{figure}[tbp]
\includegraphics[width=0.55\columnwidth, clip]{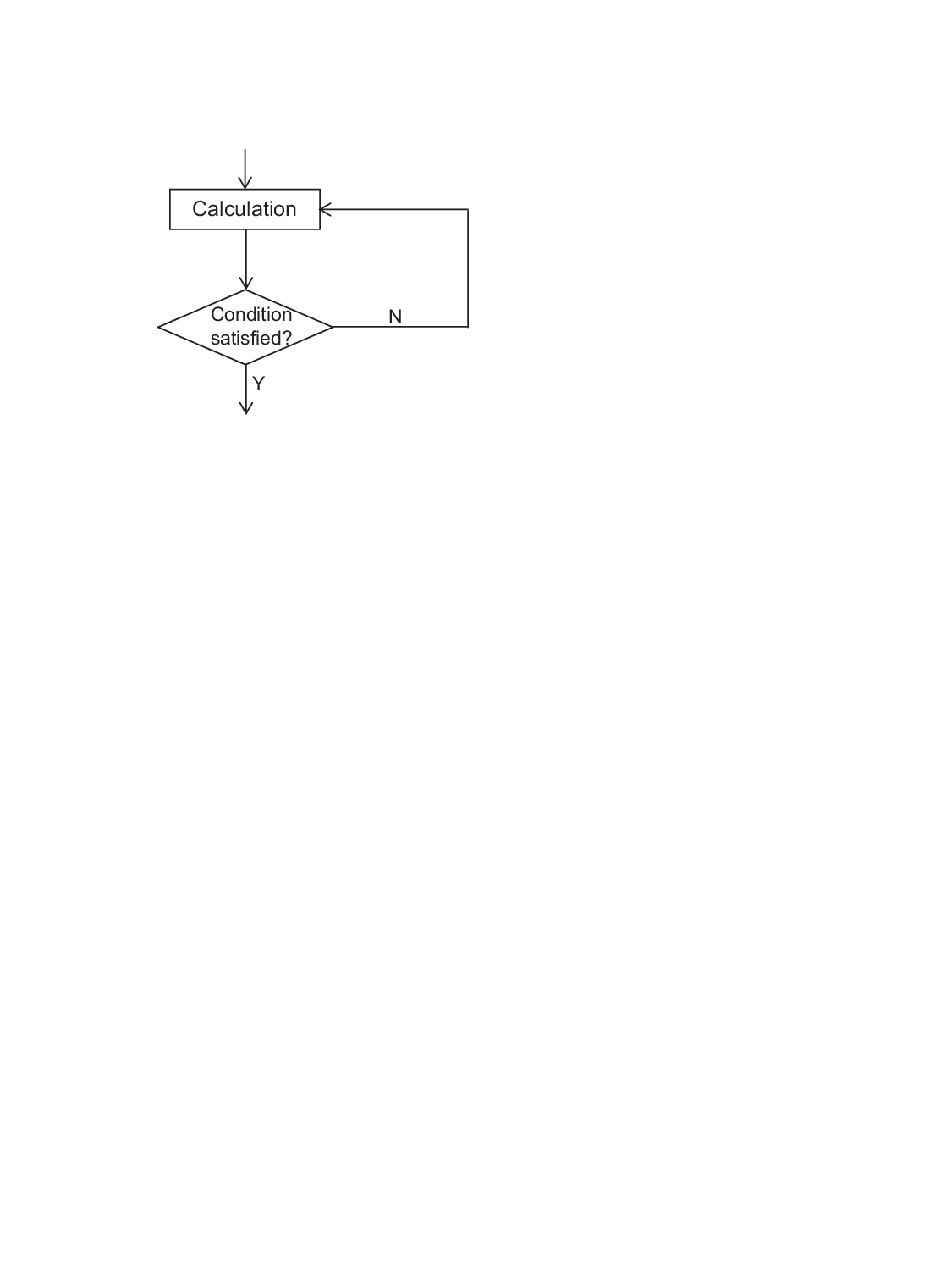}
\caption{Flowchart for one step in a multi-step computation process.}
\end{figure}

In the circuit model, a quantum computation is performed by first preparing
qubits in an initial state, then applying a series of unitary operations,
finally measuring the qubits to obtain calculation results. While in
classical computation, a problem can be solved in multiple steps, where each
step contains a computation procedure as shown in Fig.~$1$: with calculated
results of the previous step as input, one performs a calculation, then
checks if the results satisfy certain conditions; if the conditions are
satisfied, then continue calculation of the next step, otherwise, repeat the
procedure iteratively until the desired results are obtained. This process
is easy to implement in classical computation since calculated results of
each step can be copied and used repeatedly, the runtime is proportional to
the number of computation steps. In quantum computation, however, it is
difficult to realize a similar multi-step quantum computation process due to
the restriction of the no-cloning theorem~\cite{noclone1, noclone2}, which
forbids making copies of an unknown quantum state perfectly. Calculated
results encoded in an unknown quantum state cannot be used by making copies
as in classical computation. Therefore in multi-step quantum computation, if
one fails to obtain desired results of a step, one has to run the algorithm
from beginning again. This leads to the result that the runtime scales
exponentially with the number of steps of the algorithm.

We find a method to protect and reuse an unknown quantum state that encodes
calculated results of an intermediate step without copying it. Using this
method, we present a quantum algorithm for finding the ground state of a
problem Hamiltonian via multi-step quantum computation. And apply it for
efficiently solving a type of structured search problems that can be
decomposed in a more general way than that of in Refs.~\cite{struct,
structAQC}, in which the search space of the problems is reduced in
polynomial rate to the target state, while it is difficult to solve them
through the usual quantum computation process.

The idea of the algorithm is as follows: we construct an evolution path from
an initial Hamiltonian $H_{0}$ to a problem Hamiltonian $H_{P}$ by inserting
between them a sequence of intermediate Hamiltonians \{$H_{l},$ $%
l\!=\!1,\cdots ,m\!-\!1$\}, through which $H_{0}$ reaches $H_{P}$ as $%
H_{0}\rightarrow H_{1}\rightarrow \cdots \rightarrow H_{m-1}\rightarrow
H_{P}\!=\!H_{m}$. We start from the ground state $|\varphi _{0}^{(0)}\rangle
$ of $H_{0}$, and evolve it through ground states of the intermediate
Hamiltonians sequentially to reach the ground state $|\varphi
_{0}^{(m)}\rangle $ of $H_{P}$ in $m$ steps. In each step, ground state of
an intermediate Hamiltonian is obtained deterministically via quantum
resonant transitions~(QRT)~\cite{whf0, whf2}. For Hamiltonians that can be
simulated efficiently on a quantum computer, the algorithm can be run efficiently if: $i$) overlaps between ground states of any two adjacent Hamiltonians and, $ii$) energy gap between the ground and the first excited states of each Hamiltonian, are not exponentially small. The conditions can be reduced to simpler form for problems with special structures.

Compared with the algorithm in Ref.~\cite{childs}, our algorithm is flexible
in designing evolution paths. As we demonstrate later, our algorithm can
solve a type of structured search problems efficiently through a path even
when the usual AQC algorithm, which has the same efficiency as the algorithm
in~\cite{childs}, fails following the same path.

\section{The algorithm}

We describe the algorithm by using one of its steps as example. By optimizing the algorithm in~\cite{whf0, whf2}, one qubit is saved in this algorithm. It requires $\left( n+1\right) $ qubits with one probe qubit and an $n$-qubit register $R$ representing a problem of dimension $N=2^{n}$. In the $l$-th step, given the Hamiltonians $H_{l}$, $%
H_{l-1}$ and its ground state $|\varphi _{0}^{\left( l-1\right) }\rangle $
prepared on register $R$ and the corresponding eigenvalue $E_{0}^{\left(
l-1\right) }$ obtained from previous step, we aim to prepare the ground
state $|\varphi _{0}^{\left( l\right) }\rangle $ and obtain the
corresponding eigenvalue $E_{0}^{\left( l\right) }$ of $H_{l}$. The
algorithm Hamiltonian of the $l$-th step is
\begin{equation}
H^{\left( l\right) }=-\frac{1}{2}\omega \sigma _{z}\otimes
I_{N}+H_{R}^{\left( l\right) }+c\sigma _{x}\otimes I_{N},
\end{equation}%
where
\begin{equation}
H_{R}^{\left( l\right) }=\alpha _{l}|1\rangle \langle 1|\otimes
H_{l-1}+|0\rangle \langle 0|\otimes H_{l}\mathbf{,}\text{\ }l=1,2,\cdots ,m,
\end{equation}%
$I_{N}$ is the $N$-dimensional identity operator, and $\sigma _{x,\text{ }z}$
are the Pauli matrices. The first term in Eq.~($1$) is the Hamiltonian of
the probe qubit, the second term contains the Hamiltonian of the register $R$
and describes the interaction between the probe qubit and $R$, and the third
term is a perturbation. The parameter $\alpha _{l}$ is used to rescale
energy levels of $H_{l-1}$, and the ground state energy of $\alpha
_{l}H_{l-1}$ is used as a reference point to the ground state eigenvalue of $%
H_{l}$, and $c\ll 1$. We estimate the range of the ground state eigenvalue $%
E_{0}^{\left( l\right) }$ of $H_{l}$ and obtain the estimated transition
frequency range $\left[ \omega _{\min }\text{, }\omega _{\max }\right] $
between states $|\varphi _{0}^{\left( l-1\right) }\rangle $ and $|\varphi
_{0}^{\left( l\right) }\rangle $. Then discretize the frequency range into a
number of grids and use them as detection frequency of the probe qubit.
Procedures of the $l$-th step of the algorithm are as follows:

$i$) Set the probe qubit in a frequency from the frequency set, and
initialize it in its excited state $|1\rangle $ and the register $R$ in
state $|\varphi _{0}^{\left( l-1\right) }\rangle $.

$ii$) Implement the time evolution operator $U(t)=\exp \left( -iH^{\left(
l\right) }t\right) $.

$iii$) Read out the state of the probe qubit.

We repeat procedures $ii$)-$iii$) a number of times. If the measurement on
the probe qubit results in state $|1\rangle $, it indicates that register $R$
remains in state $|\varphi _{0}^{\left( l-1\right) }\rangle $, then we run
procedures $ii$)-$iii$) by setting the probe qubit in another frequency.
Otherwise if the probe qubit decays to state $|0\rangle $, it indicates a
resonant transition from state $|1\rangle |\varphi_{0}^{\left( l-1\right)
}\rangle $ to $|0\rangle |\varphi _{0}^{\left(l\right) }\rangle $. The
eigenvalue $E_{0}^{\left( l\right) }$ of $H_{l}$ can be obtained by locating
resonant transition frequency of the probe qubit that satisfies $%
E_{0}^{\left( l\right) }-\alpha _{l}E_{0}^{\left( l-1\right) }=\omega $. The
corresponding eigenvector $|\varphi _{0}^{\left( l\right) }\rangle $ can be
prepared by running the above procedure at the resonant transition frequency~%
\cite{whf0, whf2}. With $H_{l}$, $|\varphi _{0}^{\left( l\right) }\rangle $
and $E_{0}^{\left( l\right) }$, we run the algorithm for next step.
Proceeding step by step, finally we obtain the ground state of the problem
Hamiltonian. For some problems, the ground state eigenvalues of the
intermediate Hamiltonians can be calculated analytically, implementation of
the algorithm becomes easier.

As resonant transition occurs, the system is approximately in state $\sqrt{%
1-p_{0}^{\left( l\right) }}|1\rangle |\varphi _{0}^{\left( l-1\right)
}\rangle +\sqrt{p_{0}^{\left( l\right) }}|0\rangle |\varphi _{0}^{\left(
l\right) }\rangle $, where $p_{0}^{\left( l\right) }=\sin ^{2}\left(
ctd_{0}^{(l)}\right) $ is the decay probability of the probe qubit of $l$-th
step, and $d_{0}^{(l)}=\langle \varphi _{0}^{(l-1)}|\varphi
_{0}^{(l)}\rangle $. The state $|\varphi _{0}^{\left( l-1\right) }\rangle $
is protected in this entangled state. If measurement performed on the probe
results in state $|0\rangle $, it indicates the state $|\varphi _{0}^{\left(
l\right) }\rangle $ is obtained on register $R$, then we run ($l+1$)-th step
of the algorithm. Otherwise if the probe is in state $|1\rangle $, it means $%
R$ remains in state $|\varphi _{0}^{\left( l-1\right) }\rangle $, we repeat
procedures $ii$)-$iii$) by setting the probe in the resonant transition
frequency until it decays to state $|0\rangle $. By protecting calculated
results of an intermediate step in this entangled state, we do not need to
run the algorithm from beginning once it fails to obtain the desired state
in a step of the algorithm. We just repeat procedures of the step
until the desired state is obtained. With this property, desired state of
each step is obtained deterministically in polynomial time if the conditions
of the algorithm are satisfied. Here \textquotedblleft
deterministically\textquotedblright\ means that by running the procedures of a
step repeatedly, we know exactly when the desired state of the step is
obtained from the outcome of measurement on the probe qubit. The number of
times the procedures have to be repeated is proportional to $1/p_{0}^{\left(
l\right) } $. Therefore, runtime of the algorithm is proportional to $%
\sum_{l=1}^{m}1/p_{0}^{\left( l\right) }$, which scales linearly with the
number of steps of the algorithm, provided $p_{0}^{\left( l\right) }$ are
not exponentially small.

There are various ways to construct evolution paths that satisfy conditions
of the algorithm. Here we present two methods: $i$) for a system Hamiltonian
$H_{0}+V$, by writing $H=H_{0}+gV$ ($g\in \left[ 0,1\right] $) and
discretizing the parameter $g$, intermediate Hamiltonians $%
H_{l}=H_{0}+g_{l}V $ ($0=g_{0}<g_{1}<\cdots <g_{m}=1$) can be constructed~%
\cite{poulin}. The parameters $g_{l}$ can be adjusted to make $d_{0}^{(l)}$
finite; $ii$) the system Hamiltonian can be a Hamiltonian matrix,
intermediate Hamiltonian matrices can be constructed by spanning the system
Hamiltonian in a sequence of basis sets with increasing dimension, such that
the matrix $H_{l-1}$ is contained in a subspace of the following matrix $%
H_{l}$. The dimension of the basis sets can be adjusted to make $d_{0}^{(l)}$
to be finite. The path can also be constructed considering the structure of
the problem as below.

In $l$-th step, the probability of the initial state being evolved to the
state $|0\rangle |\varphi _{0}^{\left( l\right) }\rangle $ reaches maximum
at $t=\pi /(2cd_{0}^{(l)})$. Errors are introduced as the initial state
leaks to the excited states $|0\rangle |\varphi _{j}^{(l)}\rangle $ ($%
j=1,\ldots ,N-1$) with probability $p_{j}$. By assuming $%
E_{1}^{(l)}-E_{0}^{(l)}\gg cd_{0}^{(l)}$ and $\alpha
_{l}(E_{1}^{(l-1)}-E_{0}^{(l-1)})\gg cd_{0}^{(l)}$, and set the optimal
runtime $t=\pi /(2cd_{0}^{(l)})$ for convenience, we have
\begin{equation}
\sum_{j}p_{j}\leq a_{l}^{2}c^{2},
\end{equation}%
where
\begin{equation*}
a_{l}^{2}=\frac{4\left[ 1-\left( d_{0}^{(l)}\right) ^{2}\right] }{\left[
E_{1}^{(l)}-\!E_{0}^{(l)}\right] ^{2}}.
\end{equation*}%
(see Appendix A for details). If the energy gap $E_{1}^{\left(
l\right) }-E_{0}^{\left( l\right) }$ are not exponentially small, i.e.,
bounded by a polynomial function of the problem size, then $a_{l}$ is finite
and the error in the $l$-th step is bounded by $a_{l}^{2}c^{2}$. Considering
errors accumulated in all steps, the success probability of the algorithm
satisfies
\begin{equation*}
P_{\text{succ}}\geqslant \prod_{l=1}^{m}\left[ 1-a_{l}^{2}c^{2}\right]
\geqslant \left[ 1-\left( a_{\max }c\right) ^{2}\right] ^{m},
\end{equation*}%
where $a_{\max }$ is the maximum value of $a_{l}$. The coefficient $c$ can
be set such that $a_{\max }c<1/\sqrt{m}$, then $P_{\text{succ}}>1/e$ in the
asymptotic limit of $m$. The runtime of each step is proportional to $\pi
/(2cd_{0}^{(l)})$, therefore the runtime of the algorithm scales as $%
O\left(\sum_{l=1}^{m}\frac{\pi }{2(E_{1}^{(l) }-E_{0}^{(l)}) d_{0}^{(l)}}\right)$.

We need to find the accurate ground state eigenvalues of the intermediate
Hamiltonians, and do not need to know the accurate overlap $d_{0}^{(l)}$
between the ground states of two adjacent Hamiltonians, the algorithm can be
performed by using an estimated instead of the optimal runtime. These will
cause extra cost, but the scaling is the same, which is proportional to the
number of steps of the algorithm~(see Appendix A).

The time evolution operators $U(t)=\exp \left( -iH^{\left( l\right)}t\right)
$ can be implemented efficiently using Hamiltonian simulation algorithms~%
\cite{QSP, qubitization} for simulatable Hamiltonians. Our algorithm
requires performing a single-qubit measurement and resetting the probe qubit
to its excited state, such techniques have been realized in ion-trap
experiment~\cite{qubit}. We now apply the algorithm for solving a type of
structured search problems.

\section{Search problem with a special structure}

The unstructured search
problem is to find a marked item in an unsorted database of $N$ items using
an oracle that recognizes the marked item. The oracle is defined in terms of
a problem Hamiltonian $H_{P}=-|q\rangle \langle q|$, where $|q\rangle $ is
the marked state associated with the marked item. The initial Hamiltonian is
defined as $H_{0}=-|\psi _{0}\rangle \langle \psi _{0}|$, where $|\psi
_{0}\rangle =\frac{1}{\sqrt{N}}\sum_{j=0}^{N-1}|j\rangle $. We consider a
structured search problem that can be decomposed by using $m$~(in order of $%
O\left( \log N\right) $) oracles to construct a sequence of intermediate
Hamiltonians
\begin{equation}
H_{i}=\frac{N_{i}}{N}H_{0}+\left( 1-\frac{N_{i}}{N}\right) H_{P_{i}},\text{
\ \ }i=1,2,\cdots ,m-1,  \label{hamj}
\end{equation}%
where
\begin{equation}
H_{P_{i}}=-\sum_{q_{i}\in \Pi _{i}}|q_{i}\rangle \langle q_{i}|,
\end{equation}%
and $H_{m}=H_{P}$ and $\Pi _{m}$ only contains the target state $|q\rangle $%
, and $\Pi _{1}\supset \cdots \supset \Pi _{m-1}\supset \Pi _{m}$ with sizes
$N_{1}$, $\cdots $, $N_{m-1}$, $N_{m}=1$, respectively. If $N_{i}/N_{i-1}$ ($%
i=1,2,\cdots ,m$) are not exponentially small, the problem can be solved
efficiently in $m$ steps by using our algorithm.

Define $|q_{i}^{\bot }\rangle =\frac{1}{\sqrt{N-N_{i}}}\sum_{j\notin \Pi
_{i}}|j\rangle $. In basis $\left( \left\{ |q_{i}\rangle \right\} _{q_{i}\in
\Pi _{i}}\text{, }|q_{i}^{\bot }\rangle \right) $, the Hamiltonian $H_{i}$
in Eq.~($4$) can be written as
\begin{equation}
H\!_{i}\!\!=\!\!\left( \!\!%
\begin{array}{cccc}
\frac{\xi -1}{N} & \cdots & \frac{\xi -1}{N} & \frac{\left( \xi
\!-\!1\right) \!\sqrt{\!N\!-\!N_{i}}}{N} \\
\vdots & \ddots & \vdots & \vdots \\
\frac{\xi -1}{N} & \cdots & \frac{\xi -1}{N} & \frac{\left( \xi -1\right) \!%
\sqrt{N\!-\!N_{i}}}{N} \\
\frac{\left( \!\xi \!-\!1\!\right) \!\sqrt{\!N\!-\!N_{i}}}{N} & \cdots &
\frac{\left( \!\xi \!-\!1\!\right) \!\sqrt{\!N\!-\!N_{i}}}{N} & \xi ^{2}%
\end{array}%
\!\!\right) \!-\!\xi I_{N_{i}+1},
\end{equation}%
where $\xi =1-\frac{N_{i}}{N}$ and $I_{N_{i}+1}$ is the identity matrix of
dimension $N_{i}+1$. The eigenvalues of the ground and the first excited
states of $H_{i}$ are $E_{\pm }^{\left( i\right) }=\frac{-1\pm \Delta E_{i}}{%
2}$, respectively, where $\Delta E_{i}\!=\!\sqrt{\!\left( \!1\!-\!\frac{%
2N_{i}}{N}\!\right) ^{2}\!+\!4\frac{N_{i}^{2}}{N^{2}}\left( \!1\!-\!\frac{%
N_{i}}{N}\!\right) }$ is the energy gap between them and reaches minimum $%
\sqrt{11}/3\sqrt{3}\approx 0.638$ at $N_{i}/N=1/3$. Let $\mathbf{e}=\left(
1,\cdots ,1\right) ^{\text{T}}$ and $\mathbf{0}=\left( 0,\cdots ,0\right) ^{%
\text{T}}$ be $N_{i}\times 1$ vectors, respectively, there are $N_{i}-1$
degenerate eigenstates $|e_{i}^{\prime }\rangle =\left( |e_{i}^{\perp
}\rangle ,0\right) ^{\text{T}}$ of $H_{i}$ with eigenvalue $-\xi $, where $%
|e_{i}^{\perp }\rangle $ is orthogonal to $\mathbf{e}$. These eigenstates
are uncoupled from the ground and the first excited states of $H_{i}$~\cite%
{cerf}. The condition for resonant transition between states $|1\rangle
|V_{-}^{\left( i-1\right) }\rangle $ and $|0\rangle |V_{-}^{\left( i\right)
}\rangle $ is satisfied by setting $\omega =1$ and $\alpha _{i}=\left(
E_{-}^{\left( i\right) }-1\right) /E_{-}^{\left( i-1\right) }$. The ground
state of $H_{i}$ is $|V_{-}^{\left( i\right) }\rangle =x_{1}^{\left(
i\right) }\left( \mathbf{e},0\right) ^{\text{T}}+x_{2}^{\left( i\right)
}\left( \mathbf{0},1\right) ^{\text{T}}$. After normalization, the overlap
between the ground states of two adjacent intermediate Hamiltonians is
\begin{eqnarray}
d_{0}^{(i)}\! &=&\!\langle V_{-}^{\left( i\!-\!1\right) }|V_{-}^{\left(
i\right) }\rangle \!=\!\sqrt{\frac{N_{i}}{N_{i-1}}}x_{1}^{\left(
i\!-\!1\right) \ast }x_{1}^{\left( i\right) }\!+  \notag \\
&&\!\frac{N_{i\!-\!1}\!-\!N_{i}}{\sqrt{\!N_{i\!-\!1}(N\!\!-\!\!N_{i})}}%
x_{1}^{\left( i\!-\!1\right) \ast }x_{2}^{(i)}\!\!+\!\!\sqrt{\!\!\frac{%
N\!\!-\!\!N_{i\!-\!1}}{N\!\!-\!\!N_{i}}}\!x_{2}^{\left( i\!-\!1\!\right)
\ast }\!x_{2}^{\left( i\right) }.
\end{eqnarray}%
The components $x_{1}^{\left( i\right) }$ and $x_{2}^{\left( i\right) }$ are
functions of $N_{i}/N$, and $x_{1}^{\left( i\right) }$ contributes most to $%
|V_{-}^{\left( i\right) }\rangle $. If the ratio $N_{i}/N_{i-1}$ are finite,
where $N_{0}=N$, then $d_{0}^{(i)}$ are finite, the conditions of our
algorithm are satisfied. The problem can be solved efficiently through the
path in Eq.~($4$) by setting the optimal runtime in each step since $%
d_{0}^{(i)}$ can be calculated. The overlaps between the ground states of $%
H_{i}$ and $H_{P}$ is proportional to $\frac{1}{\sqrt{N_{i}}}|x_{1}^{\left(
i\right) }|$, which increase monotonically as $H_{i}$ approaching $H_{P}$.

We find some problems have the structure described above~\cite{whfuture}, as
an example, we apply our algorithm for solving the Deutsch-Jozsa
problem~(see Appendix B). Grover's algorithm cannot achieve the
same efficiency as our algorithm in solving the structured search problem by
using $m$ oracles~(see Appendix D). The unstructured search
problem has one marked item and exponential unmarked items, the ratio $1/N$
is exponentially small. It cannot be divided further, and the conditions of
our algorithm cannot be satisfied. Our algorithm has the same efficiency as
Grover's algorithm~\cite{grover} and AQC algorithm~\cite{cerf,lidar} in
solving the unstructured search problem~(see Appendix C).

\section{Comparison of the algorithm with adiabatic quantum computing}

In AQC, a system evolves from the ground state of an initial Hamiltonian to
that of the problem Hamiltonian under driving Hamiltonian varying
adiabatically from the initial Hamiltonian to the problem Hamiltonian. In
our algorithm, the ground state of the problem Hamiltonian is induced step
by step through QRT following a path from the initial Hamiltonian to the
problem Hamiltonian. The structured search problem is solved efficiently.
Applying AQC for this problem with adiabatic Hamiltonian $%
H(s)=(1-s)H_{0}+sH_{P}$, $s\in \left[ 0,1\right] $, the minimum energy gap
between the ground and the first excited states of $H(s)$ is $1/\sqrt{N}$ at
$s=1/2$, the runtime scales as $O(\sqrt{N})$.
\begin{figure}[tb]
\centering
\subfigure[][ ]{ \includegraphics[scale=0.32]{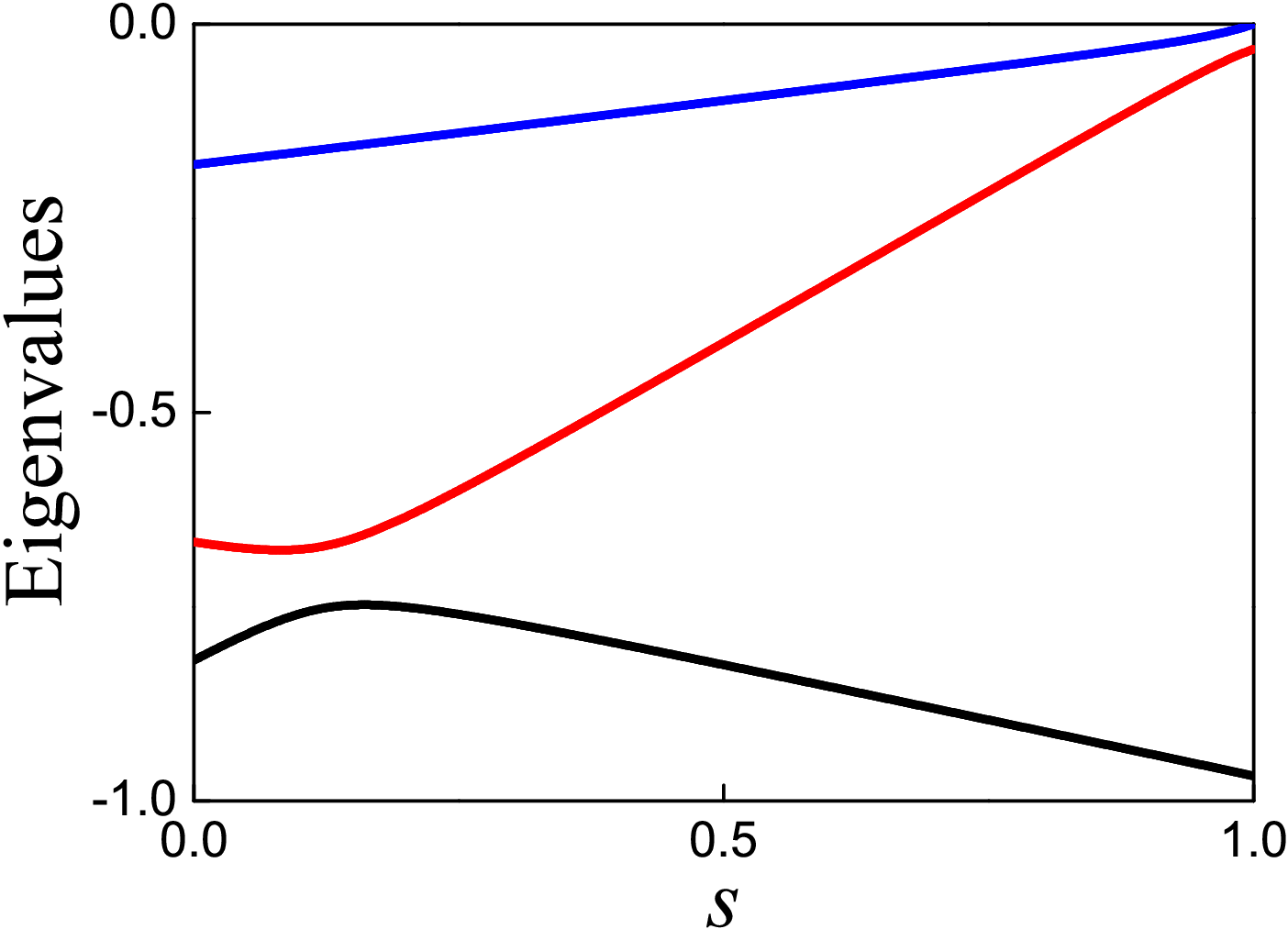}
}
\subfigure[][ ]{ \includegraphics[scale=0.32]{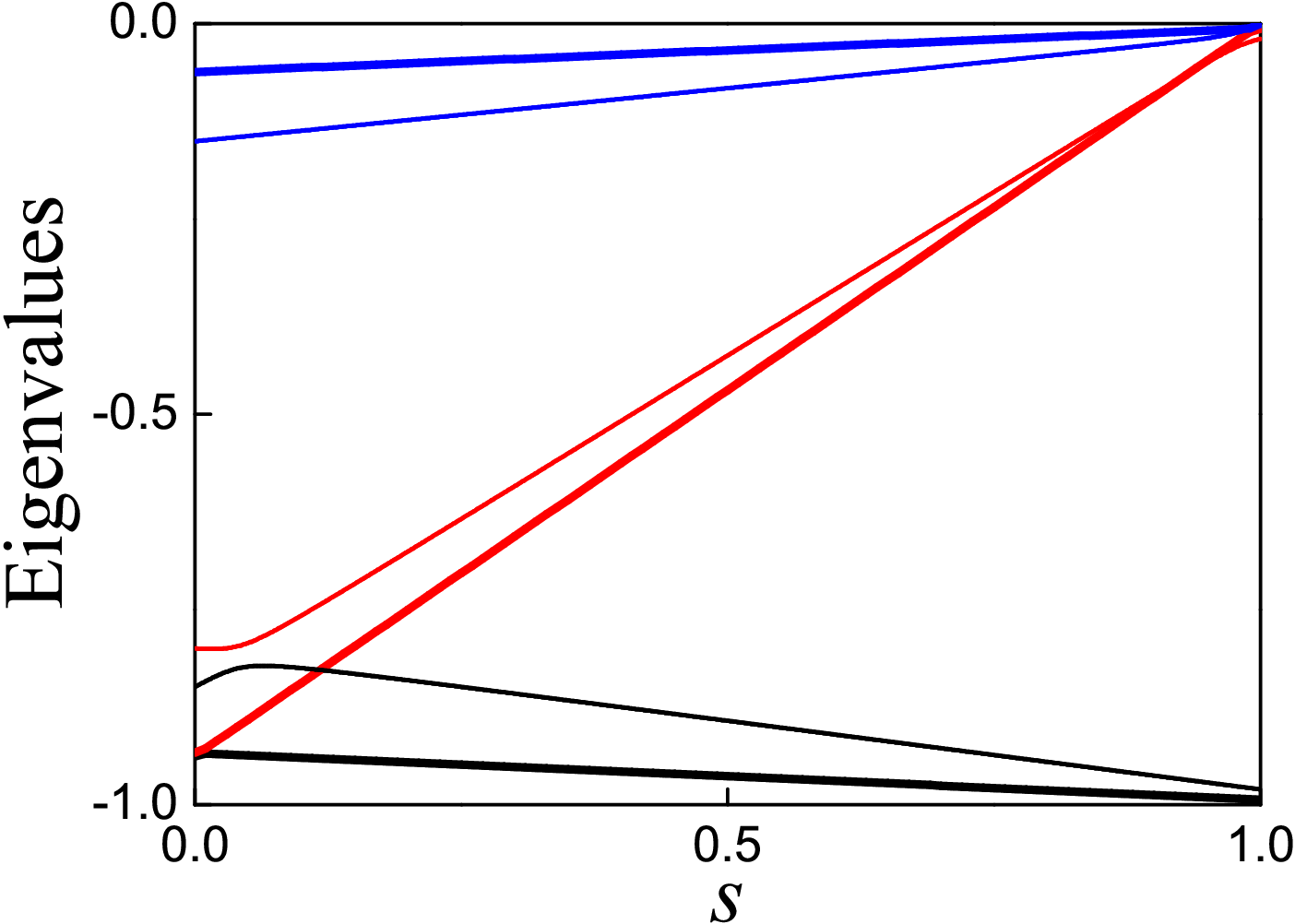}
}
\caption{ (Color online)~Eigenvalue spectrum of the adiabatic Hamiltonian $%
H_{i}^{\prime }\left( s\right) $ of an intermediate step with $%
N_{i}/N_{i-1}=1/10$ vs. the parameter $s$ by setting the parameter $N_{i}/N$
at different values. ($a$) $N_{i}/N=1/30$, ($b$) the dashed thin lines
represent the case $N_{i}/N=1/50$, and the solid thick lines represent the
case $N_{i}/N=1/150$. }
\label{Fig:gap}
\end{figure}

What about applying AQC for this problem using the same path as our
algorithm? We set $N_{i}/N_{i-1}$ to be finite, thus the conditions of our
algorithm are satisfied. Let the system evolve from the ground state of $%
H_{i-1}$ to that of $H_{i}$ under adiabatic Hamiltonian $H_{i}^{\prime
}\left( s\right) =\left( 1-s\right) H_{i-1}+sH_{i}$. We set $%
N_{i}/N_{i-1}=1/10$ and calculate eigenvalues of $H_{i}^{\prime }\left(
s\right) $ for $1<i<m$. In Fig.~\ref{Fig:gap}, we draw the energy spectrum
of $H_{i}^{\prime }\left( s\right) $ vs. $s$ where $0<s<1$ by setting $%
N_{i}/N=1/30$, $1/50$ and $1/150$, respectively. As $N_{i}/N$ becomes small,
the minimum energy gap between the ground and the first excited states
decreases quickly as $s\rightarrow 0$, and scales as $O(1/N^{2})$ at the
asymptotic limit of $N$~(see see Appendix E). Thus the usual AQC
algorithm cannot solve this problem efficiently using the path of our
algorithm. The reason for this may due to the structure coefficients $N_{i}$
of the problem are used in constructing the intermediate Hamiltonians. It
has been found that for adiabatic path constructed in linear interpolation
of two Hamiltonians, gaps can become super-exponentially small, the time for
adiabatic evolution is longer than the time required for even a classical
brute force search~\cite{AT,Hastings}. In see Appendix F, we
summarize the performance of different algorithms for solving the
structured search problems.

\section{Discussion}

We present a quantum algorithm that solves a problem through a multi-step quantum computation process in which an unknown quantum state can be protected and reused without copying it. The runtime is proportional to the number of steps of the algorithm, provided the conditions of the algorithm are satisfied. We find a type of structured
search problems can be solved efficiently by using our algorithm.
Classically one has to check the items one by one to find the target item in
solving these problems, the cost scales as the number of items $O(N)$. Our
algorithm achieves an exponential speedup over classical algorithms in
solving these problems in time that scales $O(\log N)$.

In Ref.~\cite{AT}, a jagged adiabatic path approach was proposed for running
AQC by using a sequence of Hamiltonians as stepping stones. This approach
can have the same complexity as our algorithm by using the same intermediate
Hamiltonians. It requires to project out the ground state of each
Hamiltonian to form an adiabatic path. In comparison, our algorithm is much
simpler, it needs only one ancilla qubit, and the implementation requires
only Hamiltonian simulation for which there are optimal quantum algorithms~%
\cite{QSP,qubitization} for simulatable Hamiltonians. Our algorithm of
multi-step quantum computation can be used for universal quantum computing
and developing new quantum algorithms for other problems.

\begin{acknowledgements}
We thank S.~C. Li, S. Ashhab, A. Miranowicz and F. Nori for helpful discussions. This work was supported by the Natural Science Fundamental Research Program of Shaanxi Province of China under grants 2022JM-021, the Fundamental Research Funds for the Central Universities~(Grant No.~11913291000022) and the National Key Research and Development Program of China~(Grant No.~2021YFA1000600).
\end{acknowledgements}

\vskip 3mm

\appendix

\begin{appendix}

\vskip 3mm

In the following appendix, we present the error analysis of the algorithm in the main text in Appendix A. In Appendix B, we apply our algorithm for solving the Deutsch-Jozsa problem; in Appendix C, we apply our algorithm for solving the unstructured search
problem; in Appendix D, we apply Grover's algorithm for solving the search
problem with a special structure by using a number of different oracles; in Appendix E, we study the application of the quantum adiabatic algorithm for solving the search problem with a special structure via the same evolution path as our algorithm; and in Appendix F, we summarize the performance of different algorithms for solving the search problem with a special structure.

\section{Error analysis}

In the $l$-th step of the algorithm, we are given Hamiltonians $H_{l}$, $%
H_{l-1}$ and its ground state $|\varphi _{0}^{\left( l-1\right) }\rangle $
and the ground state eigenvalue $E_{0}^{\left( l-1\right) }$ that have been
obtained from the previous $(l-1)$-th step, the goal is to prepare the
ground state $|\varphi _{0}^{\left( l\right) }\rangle $ and obtain its
corresponding eigenvalue $E_{0}^{\left( l\right) }$ of the Hamiltonian $%
H_{l} $. This can be achieved via the quantum resonant transitions~(QRT)
method by optimizing the algorithm in Ref.~\cite{whf0,whf2}. The algorithm
requires $\left( n+1\right) $ qubits with one probe qubit and an $n$-qubit
quantum register $R$ representing a Hamiltonian of dimension $N=2^{n}$. The
Hamiltonian for the $l$-th step of the algorithm is constructed as%
\begin{equation}
H^{\left( l\right) }=-\frac{1}{2}\omega \sigma _{z}\otimes
I_{N}+H_{R}^{\left( l\right) }+c\sigma _{x}\otimes I_{N},
\end{equation}%
where
\begin{equation}
H_{R}^{\left( l\right) }=\alpha _{l}|1\rangle \langle 1|\otimes
H_{l-1}+|0\rangle \langle 0|\otimes H_{l}\mathbf{,}\text{\ }l=1,2,\cdots ,m,
\end{equation}%
and $I_{N}$ is the $N$-dimensional identity operator, $\sigma _{x,\text{ }z}$
are the Pauli matrices. The first term in Eq.~(A$1$) is the Hamiltonian of
the probe qubit, the second term describes the interaction between the probe
qubit and the register $R$, and the third term is a perturbation, where $%
c\ll 1$ and $\alpha _{l}$ is an adjustable parameter. The Hamiltonians $%
H_{l-1}$ and $H_{l}$ have eigenstates $H_{l-1}|\varphi _{j}^{\left(
l-1\right) }\rangle =E_{j}^{\left( l-1\right) }|\varphi _{j}^{\left(
l-1\right) }\rangle $, and $H_{l}|\varphi _{j}^{\left( l\right) }\rangle
=E_{j}^{\left( l\right) }|\varphi _{j}^{\left( l\right) }\rangle $,
respectively, where $j=0,\cdots, N-1$.

For some problems, e.g. the search problem with a special structure in the
main text where $N_{j}$ are known, the intermediate Hamiltonians can be
solved analytically, we can calculate the ground state eigenvalues of the
intermediate Hamiltonians and the overlaps $d_{0}^{(l)}$ between ground
states of $H_{l-1}$ and $H_{l}$, and set the optimal runtime $t_{\text{opt}%
}=\pi /(2cd_{0}^{(l)})$ for running the algorithm to evolve the system to
the ground state of $H_{l}$. For these problems, the optimal runtime $t_{%
\text{opt}}$ and the eigenvalues $E_{0}^{\left( l\right) }$ are known
exactly.

In general, the intermediate Hamiltonians may not be solved analytically. We
need to find accurate value of the ground state eigenvalue of the
Hamiltonian $H_{l}$ to adjust the parameters to satisfy the resonant
transition condition in QRT, while we do not need to know accurate value of
the overlaps $d_{0}^{(l)}$. The reason for this is as follows: if we know
the overlaps $d_{0}^{(l)}$, we can set the optimal runtime $t_{\text{opt}%
}=\pi /(2cd_{0}^{(l)})$, then the resonant transition probability reaches
its maximum as the condition for QRT is satisfied, the algorithm will be
more efficient. Without knowing the exact value of $d_{0}^{(l)}$, we can set
an estimated runtime $t$ to run the QRT method, in this case, the resonant
transition probability is approximately $p_{0}^{(l)}=\sin ^{2}\left(
ctd_{0}^{(l)}\right) $. Most likely it does not reach its maximum. In this
case we just repeat the procedures of the algorithm for a few times until
the resonant transition occurs. The number of times the procedures have to
be repeated is proportional to $1/p_{0}^{(l)}$. This will add extra finite
cost of the algorithm.

Now we describe how to obtain the ground state eigenvalue $E_{0}^{\left(
l\right) }$ accurately. In order to obtain the accurate $E_{0}^{\left(
l\right) }$, we need to locate the accurate resonant transition frequency
that corresponds to the maximum transition probability of the probe qubit
for a given estimated runtime $t$. In searching for the resonant transition
frequency, we discritize an estimated probe frequency range into a number of
grids to form a frequency set. The width of the grids should be smaller than
the width of the transition peaks of the probe qubit, which is given by $%
\max \left[ cd_{0}^{(l)},1/t\right] $~\cite{whf1}. We set the probe qubit in
a frequency, and run the procedures of the algorithm for a number of times.
If a resonant transition is not observed, we try the next frequency and run
the algorithm. The cost of this process is proportional to the number of
frequency points in the discretized frequency set. Here we note that in
general, the transition frequency $\omega $ of a qubit in a digital quantum
computer is fixed. While in our algorithm, variation of the transition
frequency $\omega $ of the probe qubit can be realized equivalently by
adjusting the parameter $\alpha _{l}$, e.g. for the probe qubit frequency to
be varied in frequency range $\left[ \omega _{\min },\omega _{\max }\right] $%
, this can be equivalently realized by setting the parameter $\alpha _{l}$
in a range $\left[ \alpha _{l\max },\alpha _{l\min }\right] $, where $%
E_{0}^{\left( l\right) }-\alpha _{l\max }E_{0}^{\left( l-1\right) }=\omega
_{\min }$, and $E_{0}^{\left( l\right) }-\alpha _{l\min }E_{0}^{\left(
l-1\right) }=\omega _{\max }$.

In order to obtain the accurate value of $E_{0}^{\left( l\right) }$, we need
to locate the accurate resonant transition frequency corresponding to the
maximum transition probability of the probe qubit for a given estimated
runtime $t$. We may not obtain it in one step, most likely we have to try a
number of times back and forth around the accurate resonant transition
frequency. Suppose we are trying the frequencies of the probe qubit from
small to large values, if we pass the accurate value, we need to go back and
try another frequency. This can be done by modifying the Hamiltonian in Eq.~(A$2$) to switch the adjacent Hamiltonians $H_{l-1}$ and $H_{l}$: $%
H_{R}^{\prime \left( l\right) }=\alpha _{l}|1\rangle \langle 1|\otimes
H_{l}+|0\rangle \langle 0|\otimes H_{l-1}$, and run the procedures of the
algorithm to drive the system back to state $|\varphi _{0}^{\left(
l-1\right) }\rangle $. We can repeat this procedure while varying the
frequency of the probe qubit near the resonant transition frequency to
obtain accurate value of $E_{0}^{\left( l\right) }$. The cost is
proportional to the number of times of the procedures needs to be repeated,
which scales as $O(1/\epsilon ^{2})$ where $\epsilon $ denotes the accuracy
of $E_{0}^{\left( l\right) }$. And the cost for the case when there is a
deviation $\epsilon $ of the probe frequency from the resonant transition
frequency is analyzed at the end of this section.

In the following, we estimate errors introduced when running the algorithm
by using the accurate value of $E_{0}^{\left( l\right) }$ and the optimal
runtime $t$. Let
\begin{equation}
H_{0}^{\left( l\right) }=-\frac{1}{2}\omega \sigma _{z}\otimes
I_{N}+H_{R}^{\left( l\right) },
\end{equation}%
then the algorithm Hamiltonian $H^{\left( l\right) }$ can be written as $%
H^{\left( l\right) }=H_{0}^{\left( l\right) }+c\sigma _{x}\otimes I_{N}$.
The Hamiltonian $H_{0}^{\left( l\right) }$ is the unperturbed term and has
eigenstates%
\begin{equation}
H_{0}^{\left( l\right) }|1\rangle |\varphi _{j}^{\left( l-1\right) }\rangle
=\left( \frac{\omega }{2}\!+\!\alpha _{l}E_{j}^{(l-1)}\right) |1\rangle
|\varphi _{j}^{\left( l-1\right) }\rangle ,
\end{equation}%
and%
\begin{equation}
H_{0}^{\left( l\right) }|0\rangle |\varphi _{j}^{\left( l\right) }\rangle
=\left( \frac{-\omega }{2}\!+\!E_{j}^{(l)}\right) |0\rangle |\varphi
_{j}^{\left( l\right) }\rangle ,
\end{equation}%
where $j=0,1,\cdots ,N-1$. In the $l$-th step of the algorithm, the system
is initialized in state $|1\rangle |\varphi _{0}^{\left( l-1\right) }\rangle
$. When the resonant transition condition $E_{0}^{\left( l\right) }-\alpha
_{l}E_{0}^{\left( l-1\right) }=\omega $ is satisfied, the system is
transferred to state $|0\rangle |\varphi _{j}^{\left( l\right) }\rangle $
through QRT. Errors in the algorithm are introduced by excitations from the
initial state to the excited states $|0\rangle |\varphi _{j}^{\left(
l\right) }\rangle $\ ($j=1,\cdots ,N-1$) of the unperturbed Hamiltonian $%
H_{0}^{\left( l\right) }$ induced by the perturbation term.

For a system consists of a probe qubit coupled to a two-level system
described by the Hamiltonian in Eq.~(A$1$), the maximum transition
probability from the ground state to the excited state of the two-level
system becomes higher as the transition frequency between the two-level
system gets closer to the frequency of the probe qubit. Based on this
observation, the upper bound of the error of the algorithm, that is, the
upper bound of the transition probability from the initial state to the
excited states, can be obtained by ($i$) assuming all the excited states are
degenerate at the first excited state of the unperturbed Hamiltonian $%
H_{0}^{\left( l\right) }$, and ($ii$) without considering competition of the
transition from the initial state to the target state $|0\rangle |\varphi
_{j}^{\left( l\right) }\rangle $ through QRT, which is the ground state of
the unperturbed Hamiltonian $H_{0}^{\left( l\right) }$. With the above
analysis, the upper bound of the error in the $l$-th step of the algorithm
is the probability of the system being transferred from the initial state to
the state $|0\rangle |\varphi _{1}^{\left( l\right) }\rangle $ described by
the algorithm Hamiltonian in basis of $\left\{ |1\rangle |\varphi
_{0}^{\left( l-1\right) }\rangle ,|0\rangle |\varphi _{1}^{\left( l\right)
}\rangle \right\} $ as:
\begin{equation}
{H_{err}^{(l)}=}\left(
\begin{array}{cc}
-\frac{1}{2}\omega +E_{0}^{(l)} & c\sqrt{1-\left\vert {d_{0}^{(l)}}%
\right\vert ^{2}} \\
c\sqrt{1-\left\vert {d_{0}^{(l)}}\right\vert ^{2}} & -\frac{1}{2}\omega
+E_{1}^{(l)}%
\end{array}%
\right) ,
\end{equation}%
h{ere we have denoted $d_{j0}=\langle \varphi _{j}^{(l)}|\varphi
_{0}^{(l-1)}\rangle $} {and $d_{0}^{(l)}=d_{00}>0$}. The transition
probability from the initial state $|1\rangle |\varphi _{0}^{\left(
l-1\right) }\rangle $ to the state $|0\rangle |\varphi _{1}^{\left( l\right)
}\rangle $ is:
\begin{equation}
\frac{4c^{2}\left[ 1-\left( d_{0}^{(l)}\right) ^{2}\right] }{4c^{2}\left[
1-\left( d_{0}^{(l)}\right) ^{2}\right] +\left[ E_{1}^{(l)}-\!\alpha
_{l}E_{0}^{(l-1)}-\omega \right] ^{2}}\sin ^{2}\left[ \frac{t}{2}\sqrt{4c^{2}%
\left[ 1-\left( d_{0}^{(l)}\right) ^{2}\right] +\left[ E_{1}^{(l)}-\!\alpha
_{l}E_{0}^{(l-1)}-\omega \right] ^{2}}\right],
\end{equation}%
by using the Rabi's formula~\cite{cohen}. Therefore considering the resonant
transition condition $E_{0}^{\left( l\right) }-\alpha _{l}E_{0}^{\left(
l-1\right) }=\omega $, the upper bound of the transition probabilities from
the initial state to the excited states $|0\rangle |\varphi _{j}^{\left(
l\right) }\rangle $ ($j=1,\cdots ,N-1$) can be estimated as
\begin{equation}
\sum_{j=1}^{N-1}p_{j}\leq \frac{4c^{2}\left[ 1-\left( d_{0}^{(l)}\right) ^{2}%
\right] }{4c^{2}\left[ 1-\left( d_{0}^{(l)}\right) ^{2}\right] +\left[
E_{1}^{(l)}-\!\alpha _{l}E_{0}^{(l-1)}-\omega \right] ^{2}}<\frac{4c^{2}%
\left[ 1-\left( d_{0}^{(l)}\right) ^{2}\right] }{\left[ E_{1}^{(l)}-%
\!E_{0}^{(l)}\right] ^{2}}.
\end{equation}
We obtain the upper bound of the error introduced by transition from the
initial state to the excited states of the target Hamiltonian of the ${l}$%
-th step $H_{l}$.

We denote $a_{l}^{2}=\frac{4\left[ 1-\left( d_{0}^{(l)}\right) ^{2}\right] }{%
\left[ E_{1}^{(l)}-\!E_{0}^{(l)}\right] ^{2}}$. If the energy gap $%
E_{1}^{(l)}-E_{0}^{(l)}\gg c$ is not exponentially small, i.e., bounded by a
polynomial function of the problem size, then $a_{l}$ is a finite number and
the error in the $l$-th step is bounded by $a_{l}^{2}c^{2}$. Considering
errors accumulated in all steps of the algorithm, the success probability of
the algorithm for obtaining the ground state of the problem Hamiltonian
satisfies
\begin{equation}
P_{\text{succ}}\geqslant \prod_{l=1}^{m}\left( 1-a_{l}^{2}c^{2}\right)
\geqslant \left[ 1-\left( a_{\max }c\right) ^{2}\right] ^{m},
\end{equation}%
where $a_{\max }$ is the maximum value of $a_{l}$ for $l=1,\ldots ,m$. The
coefficient $c$ can be set such that $a_{\max }c<1/\sqrt{m}$, then $P_{\text{%
succ}}>1/e$ in the asymptotic limit of $m$.

Now we consider the cost when there is deviation $\epsilon $ of the probe
frequency from the resonant transition frequency. By ignoring the
off-resonant transition, the transition probability from the state $%
|1\rangle |\varphi _{0}^{\left( l-1\right) }\rangle $ to $|0\rangle |\varphi
_{0}^{\left( l\right) }\rangle $ is $\sin ^{2}\left( \frac{\Omega _{0}}{2}%
t\right) \frac{\left( 2cd_{0}^{(l)}\right) ^{2}}{\Omega _{0}^{2}}$ where $%
\Omega _{0}=\sqrt{\left( 2cd_{0}^{(l)}\right) ^{2}+\epsilon ^{2}}$. The
transition probability is finite if $\epsilon <2cd_{0}^{(l)}$. The number of
times that this procedure needs to be repeated scales as $O(1/\epsilon ^{2})$%
, and the cost is finite as long as $d_{0}^{(l)}$ and $c$ are not
exponentially small.

\section{Application of the algorithm for the Deutsch-Jozsa problem}

In the context of our algorithm, the Deutsch-Jozsa problem is described as:
given a function $f:\left\{ 0,1\right\} ^{n}\longmapsto \left\{ 0,-1\right\}
$ which is either constant for all values of $x$ or balanced, i.e., $f\left(
x\right) =0$ on half the inputs and $f\left( x\right) =-1$ on the other
half. The Deutsch-Jozsa problem is to determine which type the function is.
Classically, the problem requires~($2^{n-1}+1$) queries of the function $f$
in the worst case. The Deutsch-Jozsa algorithm solves this problem in a
single query of $f$.

Our algorithm can also be applied for solving this problem efficiently. In
the case where $f$ is balanced, the total number of $N=2^{n}$ states can be
divided into two sets by using an oracle $f$. The set with eigenvalue $-1$
has $N_{1}=2^{n-1}$ items, thus the ratio $N_{1}/N=1/2$. To apply our
algorithm, we set the initial state of the problem to be $|\psi _{0}\rangle =%
\frac{1}{\sqrt{N}}\sum_{j=0}^{N-1}|j\rangle $, the overlap between the
initial state and the target state $|\psi _{1}\rangle =\frac{1}{\sqrt{N_{1}}}%
\sum_{x,f(x)=-1}|x\rangle $ is $1/\sqrt{2}$, therefore our algorithm can
determine whether the function is a balanced function in time $t=\pi /(\sqrt{%
2}c)$, when observing decay of the probe qubit with probability one. On the
other hand, if the resonant transition is not observed with probability one
at time $t=\pi /(\sqrt{2}c)$, we know the function is constant. In this
case, if the constant is $0$, there will be no resonant transition; if the
constant is $-1$, then the resonant transition probability reaches one at
runtime $t=\pi /(2c)$. These cases can be easily distinguished in our
algorithm, therefore we can find out whether the function $f$ is constant or
balanced.

\section{Application of the algorithm for the unstructured search problem}

The unstructured search problem is to find a marked item in an unsorted
database of $N$ items using an oracle that recognizes the marked item. The
oracle is defined in terms of a problem Hamiltonian%
\begin{equation}
H_{P}=-|q\rangle \langle q|,
\end{equation}%
where $|q\rangle $ is the marked state associated with the marked item. The
initial Hamiltonian is defined as
\begin{equation}
H_{0}=-|\psi _{0}\rangle \langle \psi _{0}|,
\end{equation}%
where $|\psi _{0}\rangle =\frac{1}{\sqrt{N}}\sum_{j=0}^{N-1}|j\rangle $. We
can solve the unstructured search problem by using the QRT method~\cite%
{whf0,whf2} directly.

We set the Hamiltonian of the algorithm as%
\begin{equation}
H=-\frac{1}{2}\omega \sigma _{z}\otimes I_{N}+H_{R}+c\sigma _{x}\otimes
I_{N},
\end{equation}%
where
\begin{equation}
H_{R}=2|1\rangle \langle 1|\otimes H_{0}+|0\rangle \langle 0|\otimes H_{P}%
\mathbf{,}
\end{equation}%
and the coupling coefficient $c\ll 1$. We set the initial state of the
circuit as $|1\rangle |\psi _{0}\rangle $, which is the eigenstate of $H_{R}$
with eigenvalue $-2$. The transition frequency of the probe qubit is set as $%
\omega =1$, such that the resonant transition condition between states $%
|1\rangle |\psi _{0}\rangle $ and $|0\rangle |q\rangle $ is satisfied. The
overlap between the initial state $|\psi _{0}\rangle $ and the target state $%
|q\rangle $ is $d_{0}=\frac{1}{\sqrt{N}}$. By applying the QRT method, the
decay probability of the probe qubit is $p_{0}=\sin ^{2}\left(
ctd_{0}\right) $, which reaches its maximum $1$ at $t=\frac{\pi }{2cd_{0}}$.
Therefore the system can evolve to the target state in runtime scales as $O(%
\sqrt{N})$. In the case of multiple marked items in a database, the problem
Hamiltonian is defined as $H_{P}=-\sum_{q_{j}\in \Pi }|q_{j}\rangle \langle
q_{j}|$, where $\Pi $ is a set of marked states associated with the marked
items of size $N_{q}$. The overlap between the initial state $|\psi
_{0}\rangle $\ and the target state $\frac{1}{\sqrt{N_{q}}}\sum_{q_{j}\in
\Pi }|q_{j}\rangle |q_{j}\rangle $ is $\sqrt{N_{q}/N}$. By applying the QRT
method, the system can evolve to the target state in runtime scales as $O(%
\sqrt{N/N_{q}})$. In both cases, the runtime of our algorithm is the same as
that of Grover's algorithm~\cite{grover} and the adiabatic quantum
computing~(AQC) algorithm~\cite{cerf,lidar}.

\section{Solving the search problem with a special structure via Grover's
algorithm}

In the main manuscript, we have shown that the search problem with a special
structure can be solved efficiently by using the QRT method in $m$ steps,
where the problem can be decomposed by $m$ different oracles and $m$ scales
in order of $O\left( \log N\right) $. We have also shown in the above
section that by using one oracle, the unstructured search problem can be
solved by the QRT method with the same efficiency as that of the Grover's
algorithm. Then a natural question is: can the structured search problem be
solved efficiently by using the Grover's algorithm with $m$ different
oracles? We analyze this problem as follows:

The structured search problem can be decomposed by $m$ different oracles
that are defined in terms of $m$ Hamiltonians
\begin{equation}
H_{P_{j}}=-\sum_{q_{j}\in \Pi _{j}}|q_{j}\rangle \langle q_{j}|,
\end{equation}%
and $H_{P_{m}}=H_{P}=-|q\rangle \langle q|$, where $|q\rangle $ is the
marked state associated with the marked item, and the set $\Pi _{m}$ only
contains the target state $|q\rangle $, and $\Pi _{1}\supset \cdots \supset
\Pi _{m-1}\supset \Pi _{m}$, with size $N_{1}$, $\cdots $, $N_{m-1}$, $%
N_{m}=1$, respectively. Here $N_{j}$ are known and $N_{j}/N_{j-1}$ ($%
j=1,2,\cdots ,m$) are not exponentially small. The corresponding oracles are
$O_{1}$, $\cdots $, $O_{m}$, respectively. The initial Hamiltonian is
defined as $H_{0}=-|\psi _{0}\rangle \langle \psi _{0}|$, where $|\psi
_{0}\rangle =\frac{1}{\sqrt{N}}\sum_{k=0}^{N-1}|k\rangle $.

Since $N_{j}$ are already known, the fixed-point search can be performed by
using Grover's algorithm. In the first iteration, the Grover operator is
constructed as:
\begin{equation}
G_{1}=\left( 2|\psi _{0}\rangle \langle \psi _{0}|-I\right) O_{1}=R_{1}O_{1},
\end{equation}%
by applying $G_{1}$ for $O\left( \sqrt{N/N_{1}}\right) $ times, we can
obtain state $|\psi _{1}\rangle =\frac{1}{\sqrt{N_{1}}}\sum_{q_{1}\in \Pi
_{1}}|q_{1}\rangle $. The oracle $O_{1}$ is queried for $O\left( \sqrt{%
N/N_{1}}\right) $ times. The Grover operator of the second iteration\ is:%
\begin{equation}
G_{2}=\left( 2|\psi _{1}\rangle \langle \psi _{1}|-I\right) O_{2}=R_{2}O_{2},
\end{equation}%
by applying $G_{2}$ for $O\left( \sqrt{N_{1}/N_{2}}\right) $ times, we
obtain state $|\psi _{2}\rangle =\frac{1}{\sqrt{N_{2}}}\sum_{q_{2}\in \Pi
_{2}}|q_{2}\rangle $. This iteration requires the oracle $O_{1}$ to be
queried for $\sqrt{N/N_{1}}\times \sqrt{N_{1}/N_{2}}=\sqrt{N/N_{2}}$ times
and the oracle $O_{2}$ for $\sqrt{N_{1}/N_{2}}$ times. By repeating this
procedure after $m$ iterations, the target state $|q\rangle $ is obtained.
The oracle $O_{1}$ is queried for $O\left( \sqrt{N}\right) $ times, and the
oracle $O_{2}$ for $O\left( \sqrt{N_{1}}\right) $ times, \ldots, etc.

Based on the above analysis, we can see that Grover's algorithm cannot
achieve the same efficiency as our algorithm in solving the structured
search problem by using $m$ oracles. We can see that in applying Grover's
algorithm for solving the problem, only one quantum computation process is
applied. While in our algorithm, we apply a multi-step quantum computation
process. In our algorithm, an oracle is used in only one step of the
algorithm to obtain the desired state of the step. By using techniques of
QRT, post-selection and measurement on the probe qubit, the desired state of
the step is obtained deterministically. The desired state is protected
through quantum entanglement and can be used repeatedly for the next step of
the algorithm. These properties of the algorithm make it achieve better
efficiency than that of Grover's algorithm in solving the structured search
problem.

\section{Solving the search problem with a special structure via quantum
adiabatic evolution}

\begin{figure}[tbp]
\begin{minipage}{0.98\linewidth}
  \centerline{\includegraphics[width=0.56\columnwidth, clip]{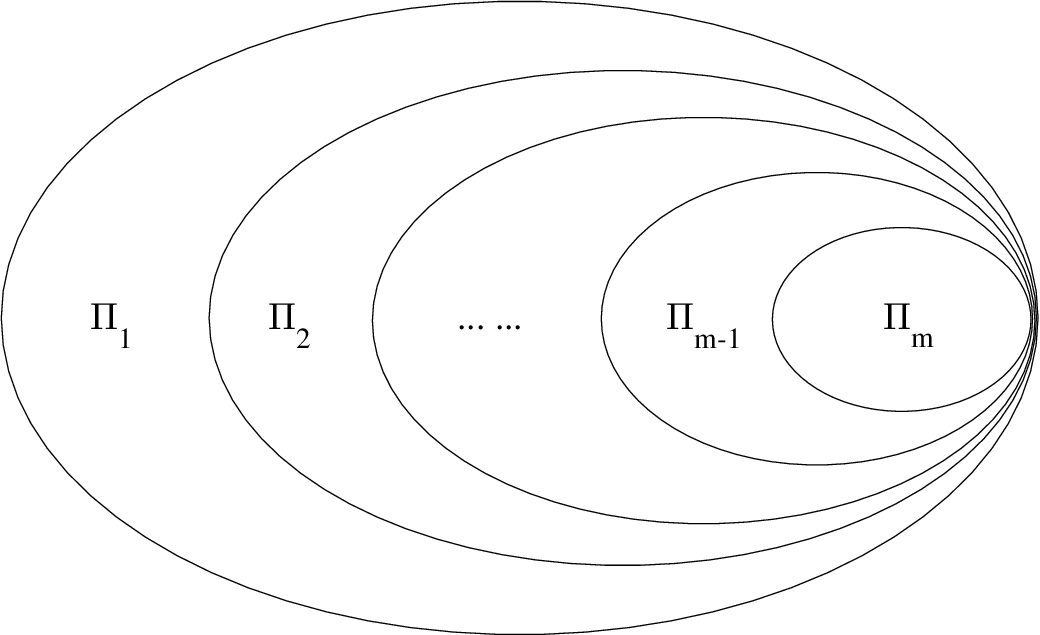}}
\end{minipage}
\caption{ The index sets $\Pi _{1}\supset \Pi _{2}\supset \cdots \supset \Pi
_{m}$.}
\label{Figure_indexSet}
\end{figure}

In the unstructured search problem, the problem Hamiltonian is $%
H_{P}=-|q\rangle \langle q|$, where $|q\rangle $ is the marked state
associated with the marked item, and the initial Hamiltonian is given by $%
H_{0}=-|\psi _{0}\rangle \langle \psi _{0}|$, where $|\psi _{0}\rangle =%
\frac{1}{\sqrt{N}}\sum_{j=0}^{N-1}|j\rangle $. The initial Hamiltonian $%
H_{0} $ has two energy levels: $-1$ with corresponding eigenstate $|\psi
_{0}\rangle $ and $0$ with $(N-1)$ eigenstates $|\psi _{0}^{\bot }\rangle $
that are orthogonal to $|\psi _{0}\rangle $. For a search problem with a
special structure that it can be decomposed by using a number of different
oracles and allows us to construct a sequence of intermediate Hamiltonians
\begin{equation}
H_{j}=\frac{N_{j}}{N}H_{0}+\left( 1-\frac{N_{j}}{N}\right) H_{P_{j}},\text{
\ }j=1,2,\cdots ,m-1,  \label{hamj}
\end{equation}%
and%
\begin{equation}
H_{P_{j}}=-\sum_{q_{j}\in \Pi _{j}}|q_{j}\rangle \langle q_{j}|,
\end{equation}%
where $\Pi _{1}\supset \cdots \supset \Pi _{m-1}$, the sizes of $\Pi _{1}$, $%
\cdots $, $\Pi _{m-1}$ are $N_{1}$, $\cdots $, $N_{m-1}$, respectively, and $%
N_{1}>\cdots >N_{m-1}$, and inserted between the initial Hamiltonian $H_{0}$
and the problem Hamiltonian $H_{P}$ (see Fig.~\ref{Figure_indexSet}). Let $%
H_{m}=H_{P}$ and the set $\Pi _{m}$ only contains the target state $%
|q\rangle $ with size $N_{m}=1$, and $\Pi _{m-1}\supset \Pi _{m}$. We
demonstrate in the main text that as long as the ratio $N_{j}/N_{j-1}$ ($%
j=1,2,\cdots ,m$) are not exponentially small, the problem can be solved
efficiently by our algorithm step by step through the evolution path $%
H_{0}\rightarrow H_{1}\rightarrow \cdots \rightarrow H_{m}=H_{P}$ as
constructed above. We start from the ground state $|\varphi
_{0}^{(0)}\rangle $ of $H_{0}$, and evolve it through ground states of the
intermediate Hamiltonians sequentially using the QRT method, finally reach
the ground state $|\varphi _{0}^{(m)}\rangle $ of $H_{P}$ in $m$ steps. An
interesting question is: will this problem be solved efficiently through
quantum adiabatic evolution using the same evolution path of our algorithm?
In the following, we study the quantum adiabatic evolution from the ground
state of the initial Hamiltonian $H_{0}$ to that of the problem Hamiltonian $%
H_{P}$ through the evolution path of our algorithm step by step. In each
step, the system is evolved adiabatically from the ground state of the
Hamiltonian $H_{j-1}$ to that of its next neighbor $H_{j}$, finally reach
the ground state of the problem Hamiltonian $H_{P}$. Note that here we apply
the usual quantum adiabatic evolution Hamiltonian $H_{j}^{\prime }\left(
s\right) =\left( 1-s\right) H_{j-1}+sH_{j}$, $s\in \left[ 0,1\right] $ in
each step.

\subsection{Eigen-problem of the intermediate Hamiltonians}

First we solve the eigen-problem of an intermediate Hamiltonian $H_{j}$.
Define
\begin{equation}
|\psi _{0}\rangle =\frac{1}{\sqrt{N}}\sum_{j=0}^{N-1}|j\rangle =\frac{1}{%
\sqrt{N}}\sum_{q_{j}\in \Pi _{j}}|q_{j}\rangle +\sqrt{\frac{N-N_{j}}{N}}%
|q_{j}^{\bot }\rangle .
\end{equation}%
where%
\begin{equation}
|q_{j}^{\bot }\rangle =\frac{1}{\sqrt{N-N_{j}}}\sum_{k\notin \Pi
_{j}}|k\rangle .
\end{equation}%
Then in the basis $\left( \left\{ |q_{j}\rangle \right\} _{q_{j}\in \Pi _{j}}%
\text{, }|q_{j}^{\bot }\rangle \right) $ we have%
\begin{eqnarray}
H\!_{0}\!\! &=&-|\psi _{0}\rangle \langle \psi _{0}|  \notag \\
&=&-\!\!\left( \!\!%
\begin{array}{cccc}
\frac{1}{N} & \cdots & \frac{1}{N} & \frac{\sqrt{\!N\!-\!N_{j}}}{N} \\
\vdots & \ddots & \vdots & \vdots \\
\frac{1}{N} & \cdots & \frac{1}{N} & \frac{\sqrt{N\!-\!N_{j}}}{N} \\
\frac{\sqrt{\!N\!-\!N_{j}}}{N} & \cdots & \frac{\sqrt{\!N\!-\!N_{j}}}{N} &
\frac{\!N\!-\!N_{j}\!}{N}\!%
\end{array}%
\!\!\right) ,
\end{eqnarray}%
and
\begin{eqnarray}
H_{P_{j}} &=&-\sum_{q_{j}\in \Pi _{j}}|q_{j}\rangle \langle q_{j}|  \notag \\
&=&-\!\!\left( \!\!%
\begin{array}{cccc}
1 & \cdots & 0 & 0 \\
\vdots & \ddots & \vdots & \vdots \\
0 & \cdots & 1 & 0 \\
0 & \cdots & 0 & 0%
\end{array}%
\!\!\right) .
\end{eqnarray}%
Then
\begin{eqnarray}
H_{j} &=&\frac{N_{j}}{N}H_{0}+\left( 1-\frac{N_{j}}{N}\right) H_{P_{j}}
\notag \\
&=&-\!\!\left( \!\!%
\begin{array}{ccccc}
\frac{N_{j}}{N^{2}}+1-\frac{N_{j}}{N} & \frac{N_{j}}{N^{2}} & \cdots & \frac{%
N_{j}}{N^{2}} & \frac{N_{j}\sqrt{\!N\!-\!N_{j}}}{N^{2}} \\
\frac{N_{j}}{N^{2}} & \frac{N_{j}}{N^{2}}+1-\frac{N_{j}}{N} & \cdots & \vdots
& \vdots \\
\vdots & \vdots & \ddots &  &  \\
\frac{N_{j}}{N^{2}} & \frac{N_{j}}{N^{2}} & \cdots & \frac{N_{j}}{N^{2}}+1-%
\frac{N_{j}}{N} & \frac{N_{j}\sqrt{N\!-\!N_{j}}}{N^{2}} \\
\frac{N_{j}\sqrt{\!N\!-\!N_{j}}}{N^{2}} & \frac{N_{j}\sqrt{\!N\!-\!N_{j}}}{%
N^{2}} & \cdots & \frac{N_{j}\sqrt{\!N\!-\!N_{j}}}{N^{2}} & \frac{%
N_{j}\left( N\!-\!N_{j}\right) \!\!}{N^{2}}%
\end{array}%
\!\!\right) .
\end{eqnarray}

Let
\begin{equation}
\mathbf{e}=\left(
\begin{array}{c}
1 \\
\vdots \\
1 \\
1%
\end{array}%
\right) _{(N_{j}+1)\times 1},\qquad \mathbf{e}_{n}=\left(
\begin{array}{c}
0 \\
\vdots \\
0 \\
1%
\end{array}%
\right) _{(N_{j}+1)\times 1},  \label{Definitions_e_en}
\end{equation}%
then $H_{0}$ can be written as
\begin{eqnarray}
H_{0} &=&-\left( \frac{1}{\sqrt{N}}\mathbf{e}+\frac{\sqrt{N-N_{j}}-1}{\sqrt{N%
}}\mathbf{e}_{n}\right) \left( \frac{1}{\sqrt{N}}\mathbf{e}+\frac{\sqrt{%
N-N_{j}}-1}{\sqrt{N}}\mathbf{e}_{n}\right) ^{T}  \notag \\
&=&-\frac{1}{N}\mathbf{e}\mathbf{e}^{T}-\frac{\sqrt{N-N_{j}}-1}{N}\left(
\mathbf{e}\mathbf{e}_{n}^{T}+\mathbf{e}_{n}\mathbf{e}^{T}\right) -\frac{%
\left( \sqrt{\!N\!-\!N_{j}}-1\right) ^{2}}{N}\mathbf{e}_{n}\mathbf{e}%
_{n}^{T},
\end{eqnarray}%
and%
\begin{equation}
H_{P_{j}}=-\left( I_{n}-\mathbf{e}_{n}\mathbf{e}_{n}^{T}\right) ,
\end{equation}%
where $I_{n}$ is the $(N_{j}+1)$-dimensional identity operator.

Thus
\begin{eqnarray}
H_{j} &=&\frac{N_{j}}{N}H_{0}+\left( 1-\frac{N_{j}}{N}\right) H_{P_{j}}
\notag \\
&=&\!-\frac{N_{j}}{N}\!\left[\! \frac{1}{N}\mathbf{e}\mathbf{e}^{T}\!+\!%
\frac{\sqrt{N\!-\!N_{j}}\!-\!1}{N}\left(\! \mathbf{e} \mathbf{e}_{n}^{T}\!+\!%
\mathbf{e}_{n}\mathbf{e}^{T}\!\right) \!+\!\frac{\left(\! \sqrt{\!N\!-\!N_{j}%
}\!-\!1\!\right) ^{2}}{N}\mathbf{e}_{n}\mathbf{e}_{n}^{T} \!\right]
\!\!-\!\!\left(\! 1\!-\!\frac{N_{j}}{N} \!\right) \left(\! I_{n}\!-\!\mathbf{%
e}_{n}\mathbf{e}_{n}^{T} \!\right)  \notag \\
&=&-\frac{N_{j}}{N^{2}}\mathbf{e}\mathbf{e}^{T}\!-\!\frac{N_{j}\left(\!
\sqrt{N-N_{j}}\!-\!1 \!\right) }{N^{2}}\left( \mathbf{e} \mathbf{e}%
_{n}^{T}\!+\!\mathbf{e}_{n}\mathbf{e}^{T}\right) \!+\!\left[\! \left(\!
1\!-\!\frac{N_{j}}{N}\!\right) \!-\!\frac{N_{j}\left(\! \sqrt{\!N\!-\!N_{j}}%
\!-\!1\!\right) ^{2}}{N^{2}}\!\right] \mathbf{e}_{n}\mathbf{e}_{n}^{T}
\notag \\
&&-\left( 1-\frac{N_{j}}{N}\right) I_{n}  \notag \\
&\equiv &\alpha \mathbf{e}\mathbf{e}^{T}+\beta \left( \mathbf{e} \mathbf{e}%
_{n}^{T}+\mathbf{e}_{n}\mathbf{e}^{T}\right) +\left( \gamma -2\beta \right)
\mathbf{e}_{n}\mathbf{e}_{n}^{T}-\left( 1-\frac{N_{j}}{N}\right) I_{n},
\label{formula_intermediateHj}
\end{eqnarray}%
where $\alpha =-\frac{N_{j}}{N^{2}}$, $\beta =-\frac{N_{j}\left( \sqrt{%
N-N_{j}}-1\right) }{N^{2}}$, $\gamma = 2\beta+ 1-\frac{N_{j}}{N} -\frac{%
N_{j}\left( \sqrt{\!N\!-\!N_{j}}-1\right) ^{2}}{N^{2}}$.
%
Please note that, with a bit abuse of notation, in the following we will
reuse the notations $\mathbf{e}$, $\mathbf{e}_n$, $\alpha$, $\beta$ and $%
\gamma$, and their dimensions and values can be determined easily from the
context.

Define $\tilde{\mathbf{e}}$ as the $N_{j}\times 1$ vector of all ones,
and we can rewrite $H_{j}$ as
\begin{equation*}
H_{j}=\alpha \mathbf{e}\mathbf{e}^{T}\!+\!\left(
\begin{array}{cc}
\mathbf{0}_{N_{j}} & \beta \tilde{\mathbf{e}} \\
\beta \tilde{\mathbf{e}}^{T} & \gamma%
\end{array}%
\right) -\left( 1-\frac{N_{j}}{N}\right) I_{n}\equiv \alpha \mathbf{e}%
\mathbf{e}^{T}+G-\left( 1-\frac{N_{j}}{N}\right) I_{n}.
\end{equation*}

($i$) Define the vector space $V=\left\{ \mathbf{e}, \mathbf{e}_{n}\right\} $
of dimension $2$. Then from Eq.~\eqref{formula_intermediateHj}, $\forall
x\in V^{\perp }$, we have $H_{j}x=-\left( 1-\frac{N_{j}}{N}\right) x$,
therefore the eigenvalues are $-\left( 1-\frac{N_{j}}{N}\right) $,
corresponding to $N_{j}-1$ eigenvectors.

($ii$) The vector space of $V$ can be spanned by vectors%
\begin{equation*}
W=\left[ \frac{1}{\sqrt{N_{j}}}\left(
\begin{array}{c}
1 \\
\vdots \\
1 \\
0%
\end{array}%
\right) ,\left(
\begin{array}{c}
0 \\
\vdots \\
0 \\
1%
\end{array}%
\right) \right] .
\end{equation*}%
It is easy to check that%
\begin{equation*}
W^{T}GW=\left(
\begin{array}{cc}
0 & \beta \sqrt{N_{j}} \\
\beta \sqrt{N_{j}} & \gamma%
\end{array}%
\right) ,
\end{equation*}%
and
\begin{equation*}
W^{T}\mathbf{e}=\left(
\begin{array}{c}
\sqrt{N_{j}} \\
1%
\end{array}%
\right) .
\end{equation*}%
Then we can verify that
\begin{eqnarray}
W^{T}H_{j}W &=&\alpha W^{T}\mathbf{e}\mathbf{e}^{T}W+W^{T}GW-\left( 1-\frac{%
N_{j}}{N}\right) I  \notag \\
&=&\left(
\begin{array}{cc}
\alpha N_{j} & \left( \alpha +\beta \right) \sqrt{N_{j}} \\
\left( \alpha +\beta \right) \sqrt{N_{j}} & \alpha +\gamma%
\end{array}%
\right) -\left( 1-\frac{N_{j}}{N}\right) I.  \label{formulaHj4SinglePoint}
\end{eqnarray}

Solving the eigen-problem of the above matrix, we can obtain the
eigenvalues:
\begin{eqnarray}
\lambda _{1,2}\! &=&\frac{1}{2}\left( \alpha N_{j}\!+\!\alpha \!+\!\gamma
\right) \pm \frac{1}{2}\sqrt{(\alpha N_{j}\!+\!\alpha \!+\!\gamma
)^{2}\!+\!4(\alpha \!+\!\beta )^{2}N_{j}\!-\!4\alpha N_{j}(\alpha
\!+\!\gamma )}\!-\!\left( 1\!-\!\frac{N_{j}}{N}\right)  \notag \\
&=&-\frac{1}{2}\pm \frac{1}{2}\sqrt{1-4\frac{N_{j}}{N}+8\left( \frac{N_{j}}{N%
}\right) ^{2}-4\left( \frac{N_{j}}{N}\right) ^{3}}.
\label{eqn:lambda4IntermediateHj}
\end{eqnarray}%
Besides these $N_{j}+1$ eigenvalues above, there are also $N-(N_{j}+1)$
degenerate eigenstates with eigenvalue $0$, and they are orthogonal to both
the vector space $V=\left\{ e,e_{n}\right\} $ and the vector space $V^{\perp
}$ of dimension $N_{j}-1$.

\subsection{Quantum adiabatic evolution in the first step}

In the first step, the system evolves adiabatically from the ground state $%
|\psi _{0}\rangle $ of $H_{0}$ to the ground state of $H_{1}=\frac{N_{1}}{N}%
H_{0}+\left( 1-\frac{N_{1}}{N}\right) H_{P_{1}}$. Define
\begin{equation}
|\psi _{0}\rangle =\frac{1}{\sqrt{N}}\sum_{j=0}^{N-1}|j\rangle =\frac{1}{%
\sqrt{N}}\sum_{q_{1}\in \Pi _{1}}|q_{1}\rangle +\sqrt{\frac{N-N_{1}}{N}}%
|q_{1}^{\bot }\rangle .
\end{equation}%
where
\begin{equation}
|q_{1}^{\bot }\rangle =\frac{1}{\sqrt{N-N_{1}}}\sum_{k\notin \Pi
_{1}}|k\rangle ,
\end{equation}%
Then in the basis $\left( \left\{ |q_{1}\rangle \right\} _{q_{1}\in \Pi _{1}}%
\text{, }|q_{1}^{\bot }\rangle \right) $ we have
\begin{eqnarray}
H\!_{0}\!\! &=&-|\psi _{0}\rangle \langle \psi _{0}|  \notag \\
&=&-\!\!\left( \!\!%
\begin{array}{cccc}
\frac{1}{N} & \cdots & \frac{1}{N} & \frac{\sqrt{\!N\!-\!N_{1}}}{N} \\
\vdots & \ddots & \vdots & \vdots \\
\frac{1}{N} & \cdots & \frac{1}{N} & \frac{\sqrt{N\!-\!N_{1}}}{N} \\
\frac{\sqrt{\!N\!-\!N_{1}}}{N} & \cdots & \frac{\sqrt{\!N\!-\!N_{1}}}{N} &
\frac{\!N\!-\!N_{1}\!}{N}\!%
\end{array}%
\!\!\right) ,
\end{eqnarray}%
and
\begin{eqnarray}
H_{P_{1}} &=&-\sum_{q_{1}\in \Pi _{1}}|q_{1}\rangle \langle q_{1}|  \notag \\
&=&-\!\!\left( \!\!%
\begin{array}{cccc}
1 & \cdots & 0 & 0 \\
\vdots & \ddots & \vdots & \vdots \\
0 & \cdots & 1 & 0 \\
0 & \cdots & 0 & 0%
\end{array}%
\!\!\right) .
\end{eqnarray}%
Then
\begin{eqnarray}
H_{1} &=&\frac{N_{1}}{N}H_{0}+\left( 1-\frac{N_{1}}{N}\right) H_{P_{1}}
\notag \\
&=&-\!\!\left( \!\!%
\begin{array}{cccc}
\frac{N_{1}}{N^{2}}+1-\frac{N_{1}}{N} & \cdots & \frac{N_{1}}{N^{2}} & \frac{%
N_{1}\sqrt{\!N\!-\!N_{1}}}{N^{2}} \\
\vdots & \ddots & \vdots & \vdots \\
\frac{N_{1}}{N^{2}} & \cdots & \frac{N_{1}}{N^{2}}+1-\frac{N_{1}}{N} & \frac{%
N_{1}\sqrt{N\!-\!N_{1}}}{N^{2}} \\
\frac{N_{1}\sqrt{\!N\!-\!N_{1}}}{N^{2}} & \cdots & \frac{N_{1}\sqrt{%
\!N\!-\!N_{1}}}{N^{2}} & \frac{N_{1}(\!N\!-\!N_{1}\!)}{N^{2}}%
\end{array}%
\!\!\right) .
\end{eqnarray}%
The adiabatic evolution Hamiltonian from $H_{0}$ to $H_{1}$ is defined as
follows:%
\begin{eqnarray}
H\left( s\right) &=&\left( 1-s\right) H_{0}+sH_{1}  \notag \\
&=&\!-(\!1\!-\!s\!)\!\!\left( \!\!%
\begin{array}{cccc}
\frac{1}{N} & \cdots & \frac{1}{N} & \frac{\sqrt{\!N\!-\!N_{1}}}{N} \\
\vdots & \ddots & \vdots & \vdots \\
\frac{1}{N} & \cdots & \frac{1}{N} & \frac{\sqrt{N\!-\!N_{1}}}{N} \\
\frac{\sqrt{\!N\!-\!N_{1}}}{N} & \cdots & \frac{\sqrt{\!N\!-\!N_{1}}}{N} &
\frac{\!N\!-\!N_{1}\!}{N}\!%
\end{array}%
\!\!\right) \!\!-\!\!s\!\!\left( \!\!%
\begin{array}{cccc}
\frac{N_{1}}{N^{2}}\!\!+\!\!1\!\!-\!\!\frac{N_{1}}{N} & \cdots & \frac{N_{1}%
}{N^{2}} & \frac{N_{1}\sqrt{\!N\!-\!N_{1}}}{N^{2}} \\
\vdots & \ddots & \vdots & \vdots \\
\frac{N_{1}}{N^{2}} & \cdots & \frac{N_{1}}{N^{2}}\!\!+\!\!1\!\!-\!\!1\frac{%
N_{1}}{N} & \frac{N_{1}\sqrt{N\!-\!N_{1}}}{N^{2}} \\
\frac{N_{1}\sqrt{\!N\!-\!N_{1}}}{N^{2}} & \cdots & \frac{N_{1}\sqrt{%
\!N\!-\!N_{1}}}{N^{2}} & \frac{N_{1}(\!N\!-\!N_{1}\!)}{N^{2}}%
\end{array}%
\!\!\right) .
\end{eqnarray}

Let
\begin{equation}
\mathbf{e}=\left(
\begin{array}{c}
1 \\
\vdots \\
1 \\
1%
\end{array}%
\right) _{(N_{1}+1)\times 1},\qquad \mathbf{e}_{n}=\left(
\begin{array}{c}
0 \\
\vdots \\
0 \\
1%
\end{array}%
\right) _{(N_{1}+1)\times 1},
\end{equation}%
then $H_{0}$ can be written as
\begin{eqnarray*}
H_{0} &=&-\left( \frac{1}{\sqrt{N}}\mathbf{e}+\frac{\sqrt{N-N_{1}}-1}{\sqrt{N%
}}\mathbf{e}_{n}\right) \left( \frac{1}{\sqrt{N}}\mathbf{e}+\frac{\sqrt{%
N-N_{1}}-1}{\sqrt{N}}\mathbf{e}_{n}\right) ^{T} \\
&=&-\frac{1}{N}\mathbf{e}\mathbf{e}^{T}-\frac{\sqrt{N-N_{1}}-1}{N}\left(
\mathbf{e}\mathbf{e}_{n}^{T}+\mathbf{e}_{n}\mathbf{e}^{T}\right) -\frac{%
\left( \sqrt{\!N\!-\!N_{1}}-1\right) ^{2}}{N}\mathbf{e}_{n}\mathbf{e}%
_{n}^{T}.
\end{eqnarray*}%
Meanwhile, we have%
\begin{equation*}
H_{P_{1}}=-(I-\mathbf{e}_{n}\mathbf{e}_{n}^{T}),
\end{equation*}%
where $I$ is the $(N_{1}+1)$-dimensional identity operator. Thus,
\begin{eqnarray}
H_{1} &=&\frac{N_{1}}{N}H_{0}+\left( 1-\frac{N_{1}}{N}\right) H_{P_{1}}
\notag \\
&=&\!-\frac{N_{1}}{N}\!\!\left[ \!\frac{1}{N}\mathbf{e}\mathbf{e}%
^{T}\!\!+\!\!\frac{\sqrt{N\!-\!N_{1}}\!\!-\!\!1}{N}\left( \!\mathbf{e}%
\mathbf{e}_{n}^{T}\!\!+\!\!\mathbf{e}_{n}\mathbf{e}^{T}\!\right) \!\!+\!\!%
\frac{\left( \!\sqrt{\!N\!-\!N_{1}}\!-\!1\!\right) ^{2}}{N}\mathbf{e}_{n}%
\mathbf{e}_{n}^{T}\!\right] \!\!+\!\!\left( \!\!1\!\!-\!\!\frac{N_{1}}{N}%
\!\!\right) \!\!{\ (\!\mathbf{e}_{n}\mathbf{e}_{n}^{T}\!\!-\!\!I\!).}
\end{eqnarray}%
The adiabatic evolution Hamiltonian can be written as
\begin{eqnarray}
H\left( s\right) &=&\left( 1-s\right) H_{0}+sH_{1}=\left( 1-s+s\frac{N_{1}}{N%
}\right) H_{0}+s\left( 1-\frac{N_{1}}{N}\right) H_{P_{1}}  \notag \\
&=&\left[ -\left( 1-s\right) -s\frac{N_{1}}{N}\right] \frac{1}{N}\mathbf{e}%
\mathbf{e}^{T}+\left[ -\left( 1-s\right) -s\frac{N_{1}}{N}\right] \frac{%
\sqrt{N-N_{1}}-1}{N}\left( \mathbf{e}\mathbf{e}_{n}^{T}+\mathbf{e}_{n}%
\mathbf{e}^{T}\right)  \notag \\
&&+\!\left\{ \!\left[ \!-\left( 1-s\right) \!-\!s\frac{N_{1}}{N}\right]
\frac{\left( \sqrt{N\!-\!N_{1}}\!-\!1\right) ^{2}}{N}\!+\!s\left( 1\!-\!%
\frac{N_{1}}{N}\right) \right\} \mathbf{e}_{n}\mathbf{e}_{n}^{T}\!-\!s\left(
1\!-\!\frac{N_{1}}{N}\right) I  \notag \\
&\equiv &\alpha \mathbf{e}\mathbf{e}^{T}+\beta \left( \mathbf{e}\mathbf{e}%
_{n}^{T}+\mathbf{e}_{n}\mathbf{e}^{T}\right) +\left( \gamma -2\beta \right)
\mathbf{e}_{n}\mathbf{e}_{n}^{T}-s\left( 1-\frac{N_{1}}{N}\right) I,
\end{eqnarray}%
where $\alpha =-\left( 1-s+s\frac{N_{1}}{N}\right) \frac{1}{N}$, $\beta
=-\left( 1-s+s\frac{N_{1}}{N}\right) \frac{\sqrt{N-N_{1}}-1}{N}$,
$\gamma =-\left( 1-s+s\frac{N_{1}}{N}\right) \frac{\left( \sqrt{N-N_{1}}%
-1\right) ^{2}}{N}+s\left( 1-\frac{N_{1}}{N}\right) +2\beta =-\left( 1-s+s%
\frac{N_{1}}{N}\right) \left( 2-\frac{1}{N}-\frac{N_{1}}{N}\right) +1$.

We have
\begin{equation*}
H\left( s\right) =\alpha \mathbf{e}\mathbf{e}^{T}+\left(
\begin{array}{cc}
0_{N_{1}} & \beta \tilde{\mathbf{e}} \\
\beta \tilde{\mathbf{e}}^{T} & \gamma%
\end{array}%
\right) -s\left( 1-\frac{N_{1}}{N}\right) I\equiv \alpha \mathbf{e}\mathbf{e}%
^{T}+G-s\left( 1-\frac{N_{1}}{N}\right) I,
\end{equation*}%
where $\tilde{\mathbf{e}}$ is an $N_{1}\times 1$ vector of all ones.
Besides the $N-(N_{1}+1)$ degenerate eigenstates with trivial eigenvalue $0$%
, the other $N_{1}+1$ eigenvalues are given as follows.

($i$) Define vector space $V=\left\{ \mathbf{e},\mathbf{e}_{n}\right\} $ of
dimension $2$. Then from Eq.~(E$22$), $\forall x\in V^{\perp }$, we have $%
H\left( s\right) x=-s\left( 1-\frac{N_{1}}{N}\right) x$. Therefore, there
are $N_{1}-1$ degenerate eigenvectors, corresponding to the eigenvalue $%
-s\left( 1-\frac{N_{1}}{N}\right) $.

($ii$) The vector space $V$ can be spanned by vectors
\begin{equation*}
W=\left[ \frac{1}{\sqrt{N_{1}}}\left(
\begin{array}{c}
1 \\
\vdots \\
1 \\
0%
\end{array}%
\right) ,\left(
\begin{array}{c}
0 \\
\vdots \\
0 \\
1%
\end{array}%
\right) \right] .
\end{equation*}%
Then we have%
\begin{equation*}
W^{T}GW=\left(
\begin{array}{cc}
0 & \beta \sqrt{N_{1}} \\
\beta \sqrt{N_{1}} & \gamma%
\end{array}%
\right) ,
\end{equation*}%
and
\begin{equation*}
W^{T} \mathbf{e} =\left(
\begin{array}{c}
\sqrt{N_{1}} \\
1%
\end{array}%
\right) .
\end{equation*}%
Then
\begin{eqnarray}
W^{T}H\left( s\right) W &=&\alpha W^{T} \mathbf{e} \mathbf{e}^{T}W+W^{T}G W
- s \left( 1-\frac{N_{1}}{N}\right) I  \notag \\
&=&\left(
\begin{array}{cc}
\alpha N_{1} & \left( \alpha +\beta \right) \sqrt{N_{1}} \\
\left( \alpha +\beta \right) \sqrt{N_{1}} & \alpha +\gamma%
\end{array}%
\right) - s \left( 1-\frac{N_{1}}{N}\right) I.
\end{eqnarray}

Solving the eigen-problem of the above matrix, we can obtain the eigenvalues
and eigenstates of $H\left( s\right) $:
\begin{equation}
\lambda _{1,2}\!=-\frac{1}{2}\pm \frac{1}{2}\sqrt{1-4\left[ s\left( 1-\frac{%
N_{1}}{N}\right) ^{2}-s^{2}\left( 1-\frac{N_{1}}{N}\right) ^{3}\right] }.
\end{equation}
The energy gap reaches its minimum $\sqrt{N_{1}/N}$ at $s=\frac{N}{2(N-N_{1})%
}$. Since $N_{1}/N$ is finite, the first step can be run efficiently.

\subsection{Quantum adiabatic evolution in a middle step}

In the following, we study the evolution of the system from the ground state
of an intermediate Hamiltonian $H_{j-1}$ to that of its next neighbor $H_{j}$%
. In the basis $\left( \left\{ |q_{j-1}\rangle \right\} _{q_{j-1}\in \Pi
_{j-1}}\text{, }|q_{j-1}^{\bot }\rangle \right) $ where%
\begin{equation}
|q_{j-1}^{\bot }\rangle =\frac{1}{\sqrt{N-N_{j-1}}}\sum_{k\notin \Pi
_{j-1}}|k\rangle ,
\end{equation}%
the state $|\psi _{0}\rangle $ can be written as%
\begin{equation}
|\psi _{0}\rangle =\frac{1}{\sqrt{N}}\sum_{q_{j-1}\in \Pi
_{j-1}}|q_{j-1}\rangle +\sqrt{\frac{N-N_{j-1}}{N}}|q_{j-1}^{\bot }\rangle .
\end{equation}%
We have
\begin{eqnarray}
H\!_{0}\!\! &=&-|\psi _{0}\rangle \langle \psi _{0}|  \notag \\
&=&-\!\!\left( \!\!%
\begin{array}{cccc}
\frac{1}{N} & \cdots & \frac{1}{N} & \frac{\sqrt{\!N\!-\!N_{j-1}}}{N} \\
\vdots & \ddots & \vdots & \vdots \\
\frac{1}{N} & \cdots & \frac{1}{N} & \frac{\sqrt{N\!-\!N_{j-1}}}{N} \\
\frac{\sqrt{\!N\!-\!N_{j-1}}}{N} & \cdots & \frac{\sqrt{\!N\!-\!N_{j-1}}}{N}
& \frac{\!N\!-\!N_{j-1}\!}{N}\!%
\end{array}%
\!\!\right) ,
\end{eqnarray}%
%
%
%
%
%
%
%
%
\begin{eqnarray}
H_{P_{j-1}} &=&-\sum_{q_{j-1}\in \Pi _{j-1}}|q_{j-1}\rangle \langle q_{j-1}|
\notag \\
&=&-\!\!\left( \!\!%
\begin{array}{cccc}
1 & \cdots & 0 & 0 \\
\vdots & \ddots & \vdots & \vdots \\
0 & \cdots & 1 & 0 \\
0 & \cdots & 0 & 0%
\end{array}%
\!\!\right) ,
\end{eqnarray}%
and
\begin{eqnarray}
H_{P_{j}} &=&-\sum_{q_{j}\in \Pi _{j}}|q_{j}\rangle \langle q_{j}|  \notag \\
&=&-\left(
\begin{array}{ccccccc}
1 & \cdots & 0 & 0 & \cdots & 0 & 0 \\
\vdots & \ddots & \vdots & \vdots & \ddots & \vdots & \vdots \\
0 & \cdots & 1 & 0 & \cdots & 0 & 0 \\
0 & \cdots & 0 & 0 & \cdots & 0 & 0 \\
\vdots & \ddots & \vdots & \vdots & \ddots & \vdots & \vdots \\
0 & \cdots & 0 & 0 & \cdots & 0 & 0 \\
0 & \cdots & 0 & 0 & \cdots & 0 & 0%
\end{array}%
\right) .\!\!\!\!
\end{eqnarray}%
The above three matrices in Eqs~(E$27$), (E$28$) and (E$29$) have dimension $%
\left( N_{j-1}+1\right) \times \left( N_{j-1}+1\right) $. In the matrix $%
H_{P_{j}}$, the top-left matrix is an identity matrix with dimension of $%
N_{j}\times N_{j}$. Then $H_{j-1}$ and $H_{j}$ matrices can be written as
\begin{eqnarray}
H_{j-1} &=&\frac{N_{j-1}}{N}H_{0}+\left( 1-\frac{N_{j-1}}{N}\right)
H_{P_{j-1}}  \notag \\
&=&\!\!-\!\!\left( \!\!%
\begin{array}{ccccccc}
\frac{N_{j\!-\!1}}{N^{2}}\!\!+\!\!1\!\!-\!\!\frac{N_{j\!-\!1}}{N} & \cdots &
\frac{N_{j\!-\!1}}{N^{2}} & \frac{N_{j\!-\!1}}{N^{2}} & \cdots & \frac{%
N_{j\!-\!1}}{N^{2}} & \frac{N_{j\!-\!1}\sqrt{\!N\!-\!N_{j\!-\!1}}}{N^{2}} \\
\vdots & \ddots & \vdots & \vdots & \ddots & \vdots & \vdots \\
\frac{N_{j\!-\!1}}{N^{2}} & \cdots & \frac{N_{j\!-\!1}}{N^{2}}%
\!\!+\!\!1\!\!-\!\!\frac{N_{j\!-\!1}}{N} & \frac{N_{j\!-\!1}}{N^{2}} & \cdots
& \frac{N_{j\!-\!1}}{N^{2}} & \frac{N_{j\!-\!1}\sqrt{\!N\!-\!N_{j\!-\!1}}}{%
N^{2}} \\
\frac{N_{j\!-\!1}}{N^{2}} & \cdots & \frac{N_{j\!-\!1}}{N^{2}} & \frac{%
N_{j\!-\!1}}{N^{2}}\!+\!1\!-\!\frac{N_{j\!-\!1}}{N} & \cdots & \frac{%
N_{j\!-\!1}}{N^{2}} & \frac{N_{j\!-\!1}\sqrt{\!N\!-\!N_{j\!-\!1}}}{N^{2}} \\
\vdots & \ddots & \vdots & \vdots & \ddots & \vdots & \vdots \\
\frac{N_{j\!-\!1}}{N^{2}} & \cdots & \frac{N_{j\!-\!1}}{N^{2}} & \frac{%
N_{j\!-\!1}}{N^{2}} & \cdots & \frac{N_{j\!-\!1}}{N^{2}}\!\!+\!\!1\!\!-\!\!%
\frac{N_{j\!-\!1}}{N} & \frac{N_{j\!-\!1}\sqrt{\!N\!-\!N_{j\!-\!1}}}{N^{2}}
\\
\frac{N_{j\!-\!1}\sqrt{\!N\!-\!N_{j\!-\!1}}}{N^{2}} & \cdots & \frac{%
N_{j\!-\!1}\sqrt{\!N\!-\!N_{j\!-\!1}}}{N^{2}} & \frac{N_{j\!-\!1}\sqrt{%
\!N\!-\!N_{j\!-\!1}}}{N^{2}} & \cdots & \frac{N_{j\!-\!1}\sqrt{%
\!N\!-\!N_{j\!-\!1}}}{N^{2}} & \frac{N_{j\!-\!1}\left(
\!N\!-\!N_{j\!-\!1}\!\right) }{N^{2}}%
\end{array}%
\!\!\right) ,
\end{eqnarray}%
and%
\begin{eqnarray}
H_{j} &=&\frac{N_{j}}{N}H_{0}+\left( 1-\frac{N_{j}}{N}\right) H_{P_{j}}
\notag \\
&=&-\left(
\begin{array}{ccccccc}
\frac{N_{j}}{N^{2}}+1-\frac{N_{j}}{N} & \cdots & \frac{N_{j}}{N^{2}} & \frac{%
N_{j}}{N^{2}} & \cdots & \frac{N_{j}}{N^{2}} & \frac{N_{j}\sqrt{%
\!N\!-\!N_{j-1}}}{N^{2}} \\
\vdots & \ddots & \vdots & \vdots & \ddots & \vdots & \vdots \\
\frac{N_{j}}{N^{2}} & \cdots & \frac{N_{j}}{N^{2}}+1-\frac{N_{j}}{N} & \frac{%
N_{j}}{N^{2}} & \cdots & \frac{N_{j}}{N^{2}} & \frac{N_{j}\sqrt{%
\!N\!-\!N_{j-1}}}{N^{2}} \\
\frac{N_{j}}{N^{2}} & \cdots & \frac{N_{j}}{N^{2}} & \frac{N_{j}}{N^{2}} &
\cdots & \frac{N_{j}}{N^{2}} & \frac{N_{j}\sqrt{\!N\!-\!N_{j-1}}}{N^{2}} \\
\vdots & \ddots & \vdots & \vdots & \ddots & \vdots & \vdots \\
\frac{N_{j}}{N^{2}} & \cdots & \frac{N_{j}}{N^{2}} & \frac{N_{j}}{N^{2}} &
\cdots & \frac{N_{j}}{N^{2}} & \frac{N_{j}\sqrt{\!N\!-\!N_{j-1}}}{N^{2}} \\
\frac{N_{j}\sqrt{\!N\!-\!N_{j-1}}}{N^{2}} & \cdots & \frac{N_{j}\sqrt{%
\!N\!-\!N_{j-1}}}{N^{2}} & \frac{N_{j}\sqrt{\!N\!-\!N_{j-1}}}{N^{2}} & \cdots
& \frac{N_{j}\sqrt{\!N\!-\!N_{j-1}}}{N^{2}} & \frac{N_{j}\left(
N-N_{j-1}\right) }{N^{2}}%
\end{array}%
\right) .
\end{eqnarray}

Let
\begin{equation}
\mathbf{e}=\left(
\begin{array}{c}
1 \\
\vdots \\
1 \\
1%
\end{array}%
\right) _{(N_{j-1}+1)\times 1},\qquad \mathbf{e}_{n}=\left(
\begin{array}{c}
0 \\
\vdots \\
0 \\
1%
\end{array}%
\right) _{(N_{j-1}+1)\times 1},
\end{equation}%
then $H_{0}$ can be written as
\begin{eqnarray}
H_{0} &=&-\left( \frac{1}{\sqrt{N}}\mathbf{e}+\frac{\sqrt{N-N_{j-1}}-1}{%
\sqrt{N}}\mathbf{e}_{n}\right) \left( \frac{1}{\sqrt{N}}\mathbf{e}+\frac{%
\sqrt{N-N_{j-1}}-1}{\sqrt{N}}\mathbf{e}_{n}\right) ^{T}  \notag \\
&=&-\frac{1}{N}\mathbf{e}\mathbf{e}^{T}-\frac{\sqrt{N-N_{j-1}}-1}{N}\left(
\mathbf{e}\mathbf{e}_{n}^{T}+\mathbf{e}_{n}\mathbf{e}^{T}\right) -\frac{%
\left( \sqrt{N-N_{j-1}}-1\right) ^{2}}{N}\mathbf{e}_{n}\mathbf{e}_{n}^{T},
\end{eqnarray}

and we have
\begin{equation}
H_{P_{j-1}}=-I_{n}+\mathbf{e}_{n}\mathbf{e}_{n}^{T},\text{ \ }%
H_{P_{j}}=\left(
\begin{array}{cc}
-I_{N_{j}} & 0 \\
0 & 0_{N_{j-1}+1-N_{j}}%
\end{array}%
\right) =-I_{n}+\mathbf{e}_{n}\mathbf{e}_{n}^{T}+\sum_{k\in \Pi
_{j-1}\setminus \Pi _{j}}\mathbf{e}_{k}\mathbf{e}_{k}^{T},
\end{equation}%
where $I_{n}$ is the identity matrix of dimension $N_{j-1}+1$, and $\mathbf{e%
}_{k}$ ($k=N_{j}+1,\cdots ,N_{j-1}$) are $(N_{j-1}+1)$-dimensional vectors
with the $k$-th element being $1$ and other elements $0$, forming the basis
for the subspace associated with the set $\Pi _{j-1}\setminus \Pi _{j}$,
that is
\begin{equation}
\mathbf{e}_{k}=\left(
\begin{array}{c}
0 \\
\vdots \\
0 \\
1 \\
0 \\
\vdots \\
0%
\end{array}%
\right) _{(N_{j-1}+1)\times 1},\qquad k=N_{j}+1,\cdots ,N_{j-1}.
\end{equation}

Thus,
\begin{eqnarray}
H_{j-1} &=&\frac{N_{j-1}}{N}H_{0}+\left( 1-\frac{N_{j-1}}{N}\right)
H_{P_{j-1}}  \notag \\
&=&-\frac{N_{j-1}}{N}\left[ \frac{1}{N}\mathbf{e}\mathbf{e}^{T}+\frac{\sqrt{%
N-N_{j-1}}-1}{N}\left( \mathbf{e}\mathbf{e}_{n}^{T}+\mathbf{e}_{n}\mathbf{e}%
^{T}\right) +\frac{\left( \sqrt{N-N_{j-1}}-1\right) ^{2}}{N}\mathbf{e}_{n}%
\mathbf{e}_{n}^{T}\right]  \notag \\
&&+\left( 1-\frac{N_{j-1}}{N}\right) \mathbf{e}_{n}\mathbf{e}_{n}^{T}-\left(
1-\frac{N_{j-1}}{N}\right) I_{n},
\end{eqnarray}%
and
\begin{eqnarray}
H_{j} &=&\frac{N_{j}}{N}H_{0}+\left( 1-\frac{N_{j}}{N}\right) H_{P_{j}}
\notag \\
&=&-\frac{N_{j}}{N}\left[ \frac{1}{N}\mathbf{e}\mathbf{e}^{T}+\frac{\sqrt{%
N-N_{j-1}}-1}{N}\left( \mathbf{e}\mathbf{e}_{n}^{T}+\mathbf{e}_{n}\mathbf{e}%
^{T}\right) +\frac{\left( \sqrt{N-N_{j-1}}-1\right) ^{2}}{N}\mathbf{e}_{n}%
\mathbf{e}_{n}^{T}\right]  \notag \\
&&+\left( 1-\frac{N_{j}}{N}\right) \left( \mathbf{e}_{n}\mathbf{e}%
_{n}^{T}+\sum_{k\in \Pi _{j-1}\setminus \Pi _{j}}\mathbf{e}_{k}\mathbf{e}%
_{k}^{T}\right) -\left( 1-\frac{N_{j}}{N}\right) I_{n}.
\end{eqnarray}%
For the last step $j=m$, we set $N_{j}/N$ to be $0$, then $H_{j}$ above
reduces to $H_{m}$. The adiabatic evolution Hamiltonian from $H_{j-1}$ to $%
H_{j}$ can be written as
\begin{eqnarray}
H\left( s\right) &=&\left( 1-s\right) H_{j-1}+sH_{j}  \notag \\
&=&\left[ -\left( 1-s\right) \frac{N_{j-1}}{N}-s\frac{N_{j}}{N}\right] \frac{%
1}{N}\mathbf{e}\mathbf{e}^{T}+\left[ -\left( 1-s\right) \frac{N_{j-1}}{N}-s%
\frac{N_{j}}{N}\right] \frac{\sqrt{N-N_{j-1}}-1}{N}\left( \mathbf{e}\mathbf{e%
}_{n}^{T}+\mathbf{e}_{n}\mathbf{e}^{T}\right)  \notag \\
&&+\left\{ \left[ -\left( 1-s\right) \frac{N_{j-1}}{N}-s\frac{N_{j}}{N}%
\right] \frac{\left( \sqrt{N-N_{j-1}}-1\right) ^{2}}{N}+\left[ -\left(
1-s\right) \frac{N_{j-1}}{N}-s\frac{N_{j}}{N}\right] +1\right\} \mathbf{e}%
_{n}\mathbf{e}_{n}^{T}  \notag \\
&&+s\left( 1-\frac{N_{j}}{N}\right) \sum_{k\in \Pi _{j-1}\setminus \Pi _{j}}%
\mathbf{e}_{k}\mathbf{e}_{k}^{T}-\left[ \left( 1-\frac{N_{j-1}}{N}\right)
(1-s)+\left( 1-\frac{N_{j}}{N}\right) s\right] I_{n}  \notag \\
&\equiv &\alpha \mathbf{e}\mathbf{e}^{T}+\beta \left( \mathbf{e}\mathbf{e}%
_{n}^{T}+\mathbf{e}_{n}\mathbf{e}^{T}\right) +\left( \gamma -2\beta \right)
\mathbf{e}_{n}\mathbf{e}_{n}^{T}+\delta \sum_{k\in \Pi _{j-1}\setminus \Pi
_{j}}\mathbf{e}_{k}\mathbf{e}_{k}^{T}+\nu I_{n}.  \label{formula_H(s)}
\end{eqnarray}%
Define $\tau =\left( 1-s\right) \frac{N_{j-1}}{N}+s\frac{N_{j}}{N}$, then $%
\alpha =-\frac{\tau }{N}$, $\beta =-\tau \frac{\sqrt{N-N_{j-1}}}{N}+\frac{%
\tau }{N}$, {\ $\gamma =1-2\tau +\frac{\tau }{N}\left( N_{j-1}+1\right) $}, $%
\nu =-1+\tau $, and $\delta =s\left( 1-\frac{N_{j}}{N}\right) $. For $s\neq
0 $, let $l=N_{j-1}-N_{j}$; otherwise, $l=0$. That is, $l=\left(
N_{j-1}-N_{j}\right) \left( 1-\delta _{0s}\right) $. We have
\begin{eqnarray}
H\left( s\right) &=&\alpha \mathbf{e}\mathbf{e}^{T}+\left(
\begin{array}{ccccccc}
0 & \cdots & 0 & 0 & \cdots & 0 & \beta \\
\vdots & \ddots & \vdots & \vdots & \ddots & \vdots & \vdots \\
0 & \cdots & 0 & 0 & \cdots & 0 & \beta \\
0 & \cdots & 0 & \delta & \cdots & 0 & \beta \\
\vdots & \ddots & \vdots & \vdots & \ddots & \vdots & \vdots \\
0 & \cdots & 0 & 0 & \cdots & \delta & \beta \\
\beta & \cdots & \beta & \beta & \cdots & \beta & \gamma%
\end{array}%
\right) +\nu I_{n}  \notag \\
&=&\alpha \mathbf{e}\mathbf{e}^{T}+\left(
\begin{array}{ccc}
\mathbf{0}_{N_{j-1}-l} & \mathbf{0} & \beta \tilde{\mathbf{e}} \\
\mathbf{0} & \delta I_{l} & \beta \hat{\mathbf{e}} \\
\beta \tilde{\mathbf{e}}^{T} & \beta \hat{\mathbf{e}}^{T} & \gamma%
\end{array}%
\right) +\nu I_{n}  \notag \\
&\equiv &\alpha \mathbf{e}\mathbf{e}^{T}+G+\nu I_{n},
\label{formulaHsforMiddle}
\end{eqnarray}%
where
\begin{equation*}
\tilde{\mathbf{e}}=\left(
\begin{array}{c}
1 \\
\vdots \\
1 \\
1%
\end{array}%
\right) _{(N_{j-1}-l)\times 1},\qquad \hat{\mathbf{e}}=\left(
\begin{array}{c}
1 \\
\vdots \\
1 \\
1%
\end{array}%
\right) _{l\times 1}.
\end{equation*}

It is easy to check that there are $N-(N_{j-1}+1)$ degenerate eigenstates
with eigenvalue $0$. The other $N_{j-1}+1$ nontrivial eigenvalues associated
with Eq.~\eqref{formulaHsforMiddle} are given in the following three parts.

($i$) Define the vector space $V=\left\{ \mathbf{e},\mathbf{e}_{n},\mathbf{e}%
_{k}|k\in \Pi _{j-1}\setminus \Pi _{j}\right\} $ whose dimension is $%
N_{j-1}-N_{j}+2=l+2$. $\forall x\in V^{\perp }$, it has the form $\left(
\begin{array}{c}
\tilde{\mathbf{e}}^{\perp } \\
\mathbf{0}%
\end{array}%
\right) $, where $\tilde{\mathbf{e}}^{\perp }$ are vectors that are
orthogonal to the $N_{j}\times 1$ vector $\tilde{\mathbf{e}}$, 
and $\mathbf{0}$ is the $(N_{j-1}-N_{j}+1)\times 1$ zero vector. 
Then from Eq.~\eqref{formula_H(s)}, we can check that $H\left( s\right)
x=\nu x$, $\forall x\in V^{\perp }$, and therefore the eigenvalues are $\nu $%
, corresponding to $N_{j-1}+1-\left( N_{j-1}-N_{j}+2\right)
=N_{j-1}-l-1=N_{j}-1$ eigenvectors.

($ii$)\ For $l>1$, let $U=span\left\{
\begin{array}{c}
0_{N_{j}\times 1} \\
\hat{\mathbf{e}}^{\perp } \\
0%
\end{array}%
\right\} $, where $\hat{\mathbf{e}}^{\perp }$ are vectors that are
orthogonal to the $l\times 1$ vector $\hat{\mathbf{e}}$. 
From Eq.~\eqref{formulaHsforMiddle}: $\forall x\in U$, we have $H\left(
s\right) x=(\delta +\nu )x$, and therefore the eigenvalues are $\delta +\nu $%
, corresponding to $N_{j-1}-N_{j}-1=l-1$ eigenvectors.

($iii$) The vector space $V$ is spanned by $U$ and the following three
vectors
\begin{equation*}
W=\left[ \frac{1}{\sqrt{N_{j-1}-l}}\left(
\begin{array}{c}
1 \\
\vdots \\
1 \\
0 \\
\vdots \\
0%
\end{array}%
\right) ,\frac{1}{\sqrt{l}}\left(
\begin{array}{c}
0 \\
\vdots \\
0 \\
1 \\
\vdots \\
1 \\
0%
\end{array}%
\right) ,\left(
\begin{array}{c}
0 \\
\vdots \\
0 \\
0 \\
\vdots \\
0 \\
1%
\end{array}%
\right) \right] .
\end{equation*}%
Then we have
\begin{equation*}
W^{T}GW=\left(
\begin{array}{ccc}
0 & 0 & \beta \sqrt{N_{j-1}-l} \\
0 & \delta & \beta \sqrt{l} \\
\beta \sqrt{N_{j-1}-l} & \beta \sqrt{l} & \gamma%
\end{array}%
\right) ,
\end{equation*}%
and
\begin{equation*}
W^{T}\mathbf{e}=\left(
\begin{array}{c}
\sqrt{N_{j-1}-l} \\
\sqrt{l} \\
1%
\end{array}%
\right) .
\end{equation*}%
Therefore,
\begin{eqnarray}
W^{T}H\left( s\right) W &=&\alpha W^{T}ee^{T}W+W^{T}GW+\nu I  \notag \\
&=&\!\!\left( \!\!%
\begin{array}{ccc}
\alpha (N_{j-1}-l) & \alpha \sqrt{l(N_{j-1}\!-\!l)} & \left( \alpha
\!\!+\!\!\beta \right) \!\sqrt{N_{j-1}\!-\!l} \\
\alpha \sqrt{l(N_{j-1}-l)} & l\alpha +\delta & \left( \alpha +\beta \right)
\sqrt{l} \\
\left( \alpha \!\!+\!\!\beta \right) \!\sqrt{N_{j-1}\!-\!l} & \left( \alpha
+\beta \right) \sqrt{l} & \alpha +\gamma%
\end{array}%
\!\!\right) +\nu I.  \label{formula_WTH(s)W3x3}
\end{eqnarray}%
Note that for $s\neq 0$ we have $N_{j-1}-l=N_{j}$.

For $s=0$,
\begin{equation}
H\left( s=0\right) =H_{j-1}=\frac{N_{j-1}}{N}H_{0}+\left( 1-\frac{N_{j-1}}{N}%
\right) H_{P_{j-1}},
\end{equation}%
and the matrix in the formula Eq.~\eqref{formula_WTH(s)W3x3} reduces to a $%
2\times 2$ matrix as given in Eq.~\eqref{formulaHj4SinglePoint}.

For $s=1$,
\begin{equation}
H\left( s=1\right) =H_{j}=\frac{N_{j}}{N}H_{0}+\left( 1-\frac{N_{j}}{N}%
\right) H_{P_{j}},
\end{equation}%
and the matrix in the formula Eq.~\eqref{formula_WTH(s)W3x3} is similar to
\begin{equation*}
\left(
\begin{array}{ccc}
l\alpha +\delta +\nu & \alpha \sqrt{lN_{j}} & \left( \alpha +\beta \right)
\sqrt{l} \\
\alpha \sqrt{lN_{j}} & \alpha N_{j}+\nu & \left( \alpha +\beta \right) \sqrt{%
N_{j}} \\
\left( \alpha +\beta \right) \sqrt{l} & \left( \alpha +\beta \right) \sqrt{%
N_{j}} & \alpha +\gamma +\nu%
\end{array}%
\right) ,
\end{equation*}%
where $\alpha =-\frac{\tau }{N}$, $\beta =-\tau \frac{\sqrt{N-N_{j-1}}}{N}+%
\frac{\tau }{N}$, $\gamma =1-2\tau +\frac{\tau }{N}\left( N_{j-1}+1\right) $%
, $\nu =-1+\tau $, $\delta =1-\tau $, $l=N_{j-1}-N_{j}$, and $\tau =\frac{%
N_{j}}{N}$. The matrix above is equivalent to
\begin{eqnarray}
&&-\frac{N_{j}}{N^{2}}\!\!\left( \!\!%
\begin{array}{ccc}
N_{j\!-\!1}\!\!-\!\!N_{j} & \sqrt{(N_{j\!-\!1}\!\!-\!\!N_{j})N_{j}} & \sqrt{%
(N_{j\!-\!1}\!\!-\!\!N_{j})(N\!\!-\!\!N_{j\!-\!1})} \\
\sqrt{N_{j}(N_{j-1}\!\!-\!\!N_{j})} & \frac{N^{2}\!\!-\!\!NN_{j}\!\!+\!%
\!N_{j}^{2}}{N_{j}} & \sqrt{N_{j}(N\!\!-\!\!N_{j-1})} \\
\sqrt{(N\!\!-\!\!N_{j-1})(N_{j-1}\!\!-\!\!N_{j})} & \sqrt{%
(\!N\!\!-\!\!N_{j\!-\!1})N_{j}} & N\!\!-\!\!N_{j-1}%
\end{array}%
\!\!\right)  \notag \\
&=&\!\!-\frac{N_{j}}{N^{2}}\!\!\!\left( \!\!%
\begin{array}{ccc}
\sqrt{N_{j\!-\!1}\!-\!N_{j}} & 0 & 0 \\
0 & \sqrt{N_{j}} & 0 \\
0 & 0 & \sqrt{N\!\!-\!\!N_{j\!-\!1}}%
\end{array}%
\!\!\right) \!\!\!\!\left( \!\!%
\begin{array}{ccc}
1 & 1 & 1 \\
1 & \frac{N^{2}\!\!-\!\!NN_{j}\!+\!N_{j}^{2}}{N_{j}^{2}} & 1 \\
1 & 1 & 1%
\end{array}%
\!\!\right) \!\!\!\!\left( \!\!\!%
\begin{array}{ccc}
\sqrt{\!N_{j\!-\!1}\!\!-\!\!N_{j}} & 0 & 0 \\
0 & \sqrt{N_{j}} & 0 \\
0 & 0 & \sqrt{\!N\!\!-\!\!N_{j\!-\!1}}%
\end{array}%
\!\!\right) .  \notag
\end{eqnarray}%
Obviously, there is an eigenvalue zero, and the other two eigenvalues are
given as
\begin{equation}
\lambda _{1,2}=-\frac{1}{2}\pm \frac{1}{2}\sqrt{%
1-4N_{j}/N+8(N_{j}/N)^{2}-4(N_{j}/N)^{3}},
\end{equation}%
which are the same as those in Eq.~\eqref{eqn:lambda4IntermediateHj}. The
energy spectrum of matrix in Eq.~\eqref{formula_WTH(s)W3x3} is a continuous
function of the adiabatic evolution parameter $s$.
\begin{figure}[tbp]
\centering
\subfigure[][ ]{ \includegraphics[scale=0.23]{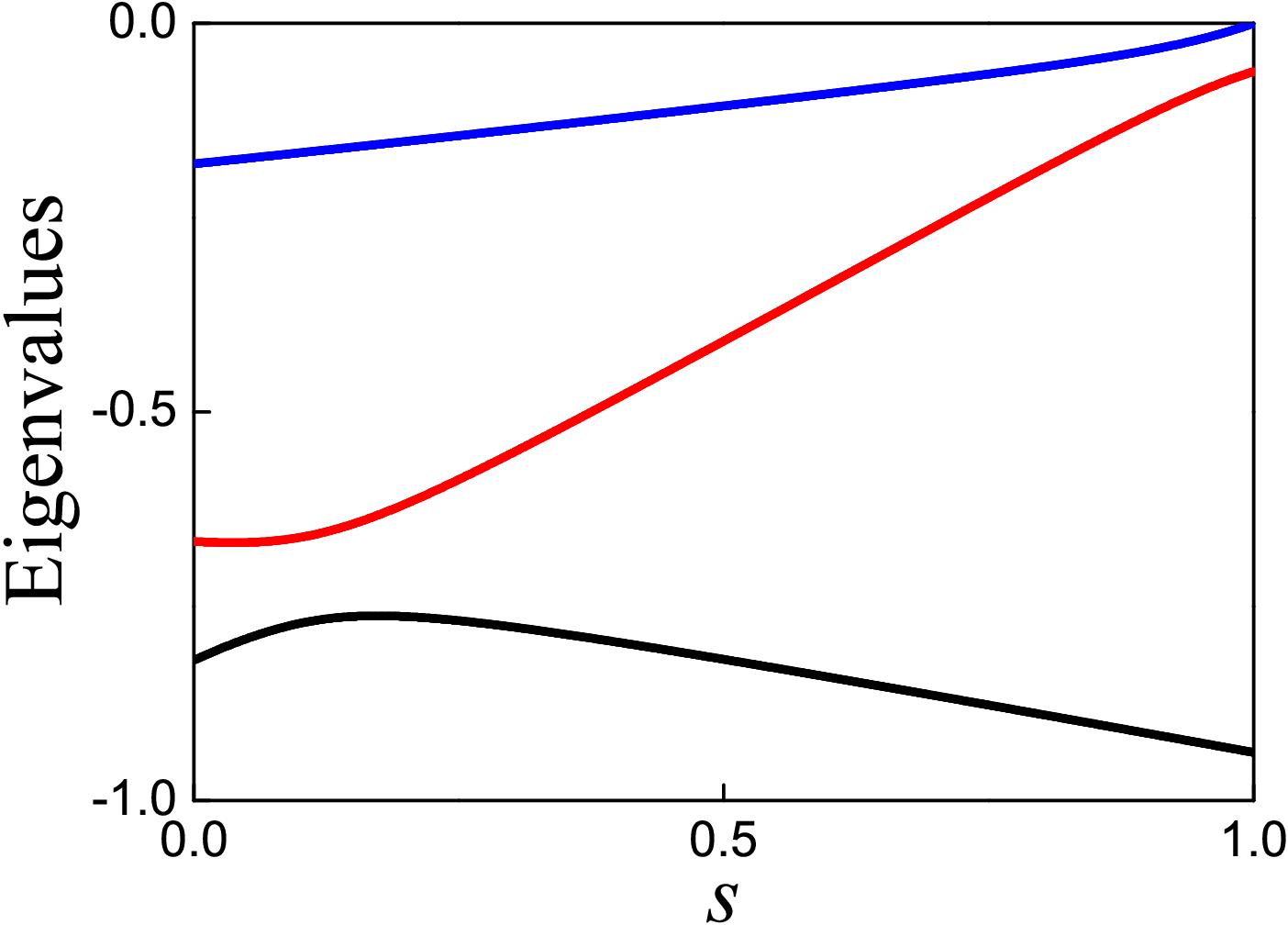}
}
\subfigure[][ ]{ \includegraphics[scale=0.23]{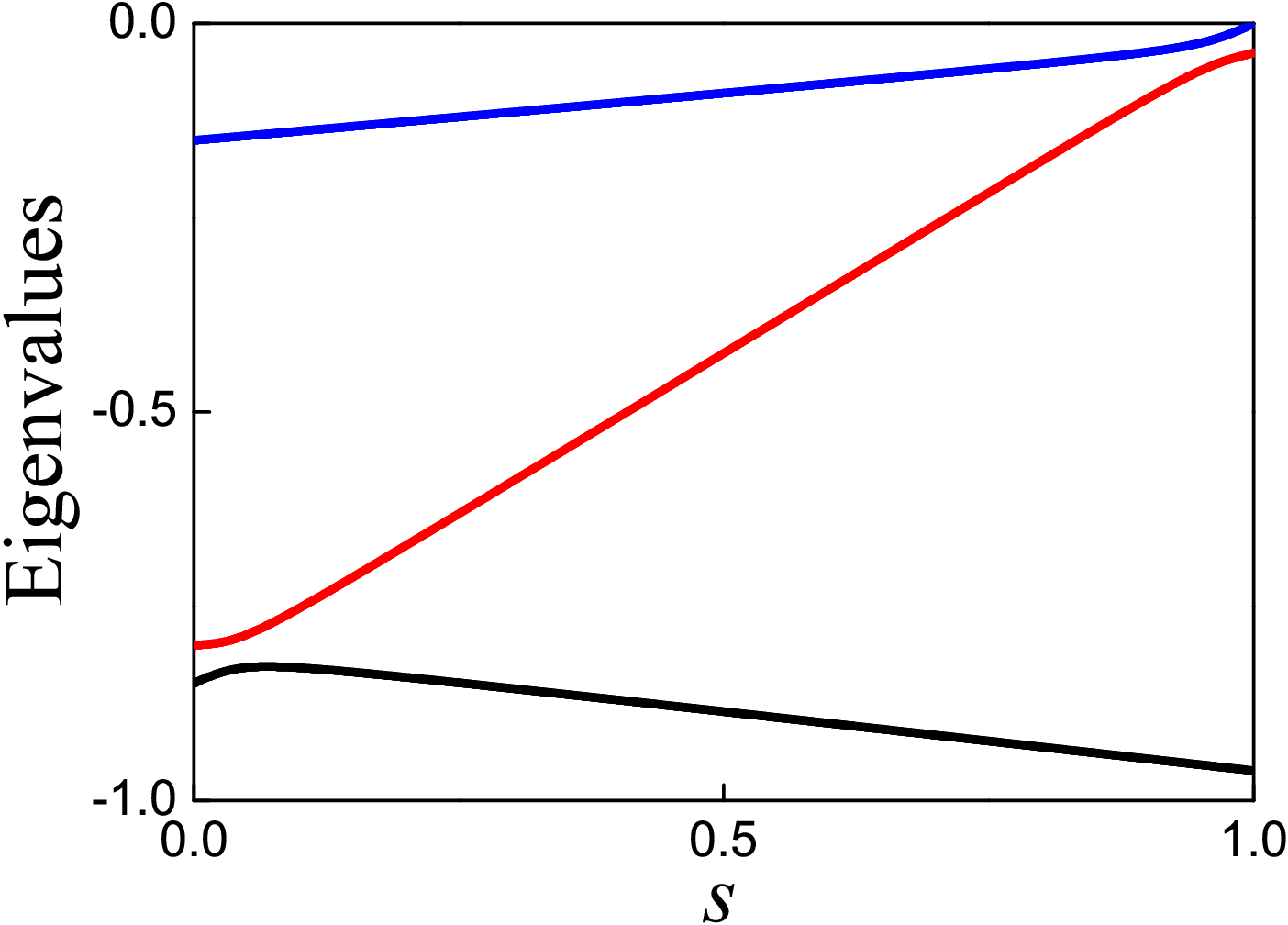}
} \subfigure[][ ]{ \includegraphics[scale=0.23]{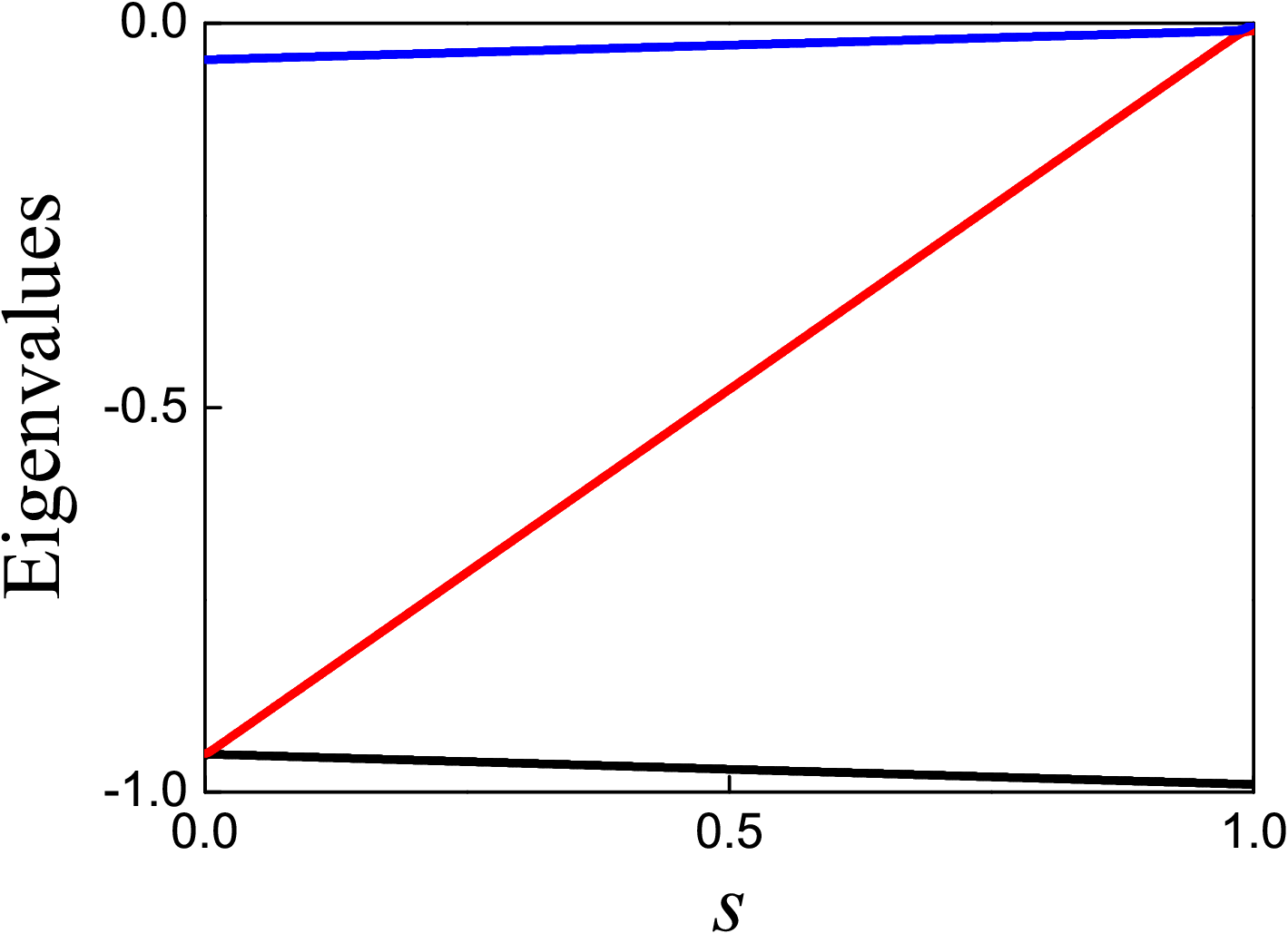}
}
\caption{ (Color online)~The eigenvalue spectrum of the adiabatic
Hamiltonian matrix of an intermediate step by setting $N_{j}/N_{j-1}=1/5$
vs. the parameter $s$ by setting $f=N_{j}/N$ at different values. ($a$) $%
f=1/15$, ($b$) $f=1/25$, ($c$) $f=1/100$. }
\end{figure}

The explicit expressions for the eigenvalues of the matrix in Eq.~%
\eqref{formula_WTH(s)W3x3} are very complicated. What we are interested is:
if the conditions for our algorithm are satisfied, will the conditions for
efficiently running the AQC algorithm be satisfied too, i.e., the minimum
energy gap between the ground and the first excited states of the adiabatic
evolution Hamiltonian be finite? Denote $g=N_{j-1}/N_{j}$ and $f=N_{j}/N$,
the matrix in Eq.~\eqref{formula_WTH(s)W3x3} can be written as
\begin{equation}
\left( \!\!%
\begin{array}{ccc}
f(f\!-\!1)\left[ s(g-1)\!-g\right] -\!1 & f^{2}\left[ s(g-1)\!-g\right]
\sqrt{g-1\!} & f^{3/2}\sqrt{1\!-\!gf}\left[ s(g-1)\!-g\right] \\
f^{2}\left[ s(g-1)\!-\!g\right] \sqrt{g\!-\!1\!} & s\left[ f^{2}(g\!-\!1)^{2}\!-\!gf\!+\!1%
\right]\!+\!gf(1\!+\!f\!-\!gf)\!\!-\!\!1 & f^{3/2}\!\sqrt{1\!-\!gf}\left[\!s(g\!-\!1)\!-\!g \!
\right]\!\sqrt{\!g\!-\!1\!} \\
f^{3/2}\sqrt{1\!-\!gf}\left[ s(g-1)\!-g\right] & f^{3/2}\sqrt{1\!-\!gf}\left[
s(g-1)\!-g\right] \sqrt{g-1\!} & f(1-gf\!)\left[ s(g-1)\!-g\right]%
\end{array}%
\!\!\right) .
\end{equation}%
In the main text, we set $N_{j}/N_{j-1}=1/10$, thus the conditions for our
algorithm are satisfied. The matrix in Eq.~\eqref{formula_WTH(s)W3x3}
becomes
\begin{equation}
\left( \!\!%
\begin{array}{ccc}
f(f\!-\!1)(9s\!-\!10)\!-\!1 & 3f^{2}(9s\!-\!10) & f^{3/2}\sqrt{1\!-\!10f}%
(9s\!-\!10) \\
3f^{2}(9s\!-\!10) & f(81fs\!-\!90f\!-\!10s\!+\!10)\!+\!s\!\!-\!\!1 & 3f^{3/2}%
\sqrt{1\!-\!10f}(9s\!-\!10) \\
f^{3/2}\sqrt{1\!-\!10f}(9s\!-\!10) & 3f^{3/2}\sqrt{1\!-\!10f}(9s\!-\!10) &
f(1\!-\!10f)(9s\!-\!10)%
\end{array}%
\!\!\right) .
\end{equation}%
We solve the eigenvalues of the above matrix for $f=1/30,1/50,1/150$,
respectively, and draw the energy spectrum vs. $s$ in the main text. We
check one more case by setting $N_{j}/N_{j-1}=1/5$, the matrix in Eq.~%
\eqref{formula_WTH(s)W3x3} becomes
\begin{equation}
\left( \!\!%
\begin{array}{ccc}
f(f\!-\!1)(4s\!-\!5)\!-\!1 & 2f^{2}(4s\!-\!5) & f^{3/2}\sqrt{1\!-5f}%
(4s\!-\!5) \\
2f^{2}(4s\!-\!5) & f(16fs\!-\!20f\!-\!5s\!+\!5)\!+\!s\!\!-\!\!1 & 2f^{3/2}%
\sqrt{1\!-\!5f}(4s\!-\!5) \\
f^{3/2}\sqrt{1\!-5f}(4s\!-\!5) & 2f^{3/2}\sqrt{1\!-\!5f}(4s\!-\!5) &
f(1\!-5f)(4s\!-\!5)%
\end{array}%
\!\!\right) .
\end{equation}%
In Fig.~$4$, we draw the energy spectrum of the above matrix as a function
of $s$ by setting $f=1/15$, $1/25$, and $1/100$, respectively. From the
figures we can see that as $f$ becomes small, the minimum energy gap between
the ground and the first excited states of the adiabatic evolution
Hamiltonians decreases quickly as $s\rightarrow 0$. Comparing with Fig.~$2$
in the main text, we can see that as $N_{j}/N_{j-1}$ becomes larger, the
minimum energy gap decreases faster as $s\rightarrow 0$. The energy gap
between the ground and the first excited states for the case where $%
N_{j}/N_{j-1}=1/5$ is in the form%
\begin{equation}
\Delta E=\frac{1}{2}\left( 1+\sqrt{1-20f+200f^{2}-500f^{3}}\right) +5f-1.
\end{equation}%
We expand the energy gap at $f\rightarrow 0$, that is, the asymptotic limit
of $N$, and find that it scales as $O(f^{2})=O(1/N^{2})$. Therefore, the
usual quantum adiabatic algorithm cannot be run efficiently by using the
same evolution path of our algorithm.

\section{Comparison of algorithms for solving the search problem with a
special structure}

In this paper, we present a quantum algorithm for preparing the ground state
of a problem Hamiltonian and apply it for solving a type of search problems
with a special structure, and compare with the AQC algorithm~(using the same
Hamiltonian evolution path as our algorithm and constructing the adiabatic
evolution Hamiltonian via linear interpolation of adjacent intermediate
Hamiltonians in each step), and the Grover's algorithm. We now summarize the
performance of these algorithms and the phase estimation algorithm~(PEA) for
solving the structured search problems.

We have shown in the above section that by using the same Hamiltonian
evolution path as our algorithm, and constructing the adiabatic evolution
Hamiltonian as a linear interpolation between two adjacent intermediate
Hamiltonians in each step as $sH_{l-1}+(1-s)H_{l}$, the runtime of the AQC
algorithm for solving the structured search problem scales as $O(N^{2})$. In
Ref.~\cite{childs}, Childs et \textit{al.} proposed a method to keep the
system in the instantaneous ground state of an adiabatic evolution
Hamiltonian by performing Zeno-like measurements. The runtime of this method
depends on the minimum energy gap between the ground and the first excited
states of the adiabatic evolution Hamiltonian. By using the continuous path
of the adiabatic evolution Hamiltonian $sH_{l-1}+(1-s)H_{l}$, the runtime of
this method has the same scaling as that of the AQC algorithm with the same
adiabatic evolution Hamiltonian. In applying the Grover's algorithm for
solving the structured search problem, the total number of all the $m$
different oracles that are called in the algorithm scales as $O(\sqrt{N})$.
The runtime of the algorithm is the same as that of solving the unstructured
search problem. The phase estimation algorithm~(PEA) is a projective method
for obtaining the ground state of a problem Hamiltonian, it projects an
initial state onto the ground state of the problem Hamiltonian with
probability proportional to the square of the overlap between them. By using
the state $|\psi _{0}\rangle =\frac{1}{\sqrt{N}}\sum_{j=0}^{N-1}|j\rangle $,
which is an uniform superposition of all the computational basis states, as
an initial guess state, the success probability of the PEA for finding the
ground state of the structured search problem is $1/N$. Therefore the
runtime of the PEA for solving the structured search problem scales as $O(N)$%
. The runtime of our algorithm for solving the structured search problem is
proportional to the number of steps $m$ which scales as $O\left( \log
N\right) $, and in each step, the runtime depends on the ratio $N_{j}/N_{j-1}
$. The algorithm is efficient as long as $N_{j}/N_{j-1}$ are not
exponentially small. In table~I, we summarize the performance of the above
algorithms for solving the search problem with a special structure.
\begin{table}[tbp]
\begin{center}
\begin{tabular}{ccp{90mm}}
\hline
Algorithms & \quad Runtime \qquad \quad & Factors determining the efficiency
of the algorithms \\ \hline
AQC & $O\left( N^{2}\right) $ & Minimum energy gap between the ground and the first
excited states of each adiabatic evolution Hamiltonian. \\
Grover & $O(\sqrt{N})$ & Number of queries of the oracles. \\
PEA & $O\left( N\right) $ & Overlap between a guess state and the ground
state of the problem Hamiltonian. \\
Our algorithm & $O\left( \log N\right) $ & Ratio $N_{j}/N_{j-1}$, which are
not exponentially small. Here $N_{j}$ represents the number of marked states
of $j$-th step. \\ \hline
\end{tabular}
\\[0pt]
\end{center}
\caption{Comparison of the performance of some algorithms for solving the
structured search problem. The AQC algorithm and the Grover's algorithm use the same Hamiltonian evolution path as our algorithm. The term $N$ represents the dimension of the search space. The performance of the method in~\protect\cite{childs} is the same as that of the AQC algorithm. }
\end{table}

These algorithms can also be applied for the general case of preparing the
ground state of a system. The runtime of the AQC algorithm scales as $%
O(1/\Delta E_{\min })$, where $\Delta E_{\min }$ is the minimum energy gap
between the ground and the first excited states of the adiabatic evolution
Hamiltonian. By using the amplitude amplification technique developed from
the Grover's algorithm, quantum algorithm can achieve quadratic speed-up
over classical algorithms in preparing the ground state of a system~\cite%
{poulin1}. The success probability of the PEA depends on the overlap between
an initial guess state and the ground state of the system. Our algorithm
requires to construct a Hamiltonian evolution path to reach the system
Hamiltonian in order to prepare the ground state of the system, the runtime
of the algorithm is proportional to summation of the evolution time of each
step, which is proportional to $1/\left( \Delta E\times d\right) $, where $%
\Delta E$ is the energy gap between the ground and the first excited states
of the target intermediate Hamiltonian and $d$ is the overlap between ground
states of two adjacent Hamiltonians of the step. The algorithm is efficient
if both $\Delta E$ of each Hamiltonian and $d$ of each step are not
exponentially small.

\end{appendix}

\end{document}